\documentclass[
reprint,
superscriptaddress,
 amsmath,amssymb,
 aps,
]{revtex4-1}

\usepackage{hyperref}
\usepackage{verbatim}
\usepackage{subcaption}
\usepackage{graphicx}
\usepackage{dcolumn}
\usepackage{bm}

\begin{document}

\title{Phase separation of stable colloidal clusters}

\author{Thomas Petersen}
\email{tapeter@mit.edu}
\affiliation{Department of Civil and Environmental Engineering, Massachusetts Institute of Technology, Cambridge, Massachusetts 02139 USA}

\author{Martin Z. Bazant}
\email{bazant@mit.edu}
\affiliation{Department of Chemical Engineering, Massachusetts Institute of Technology, Cambridge, Massachusetts 02139 USA}
\affiliation{Department of Mathematics, Massachusetts Institute of Technology, Cambridge, Massachusetts 02139 USA}

\author{Roland J.M. Pellenq}
\email{pellenq@mit.edu}
\affiliation{MultiScale Material Science for Energy and Environment, MIT-CNRS Joint Laboratory at Massachusetts Institute of Technology}
\affiliation{Department of Civil and Environmental Engineering, Massachusetts Institute of Technology, Cambridge, Massachusetts 02139 USA}

\author{Franz-Josef Ulm}
\email{ulm@mit.edu}
\altaffiliation{Corresponding Author}
\affiliation{Department of Civil and Environmental Engineering, Massachusetts Institute of Technology, Cambridge, Massachusetts 02139 USA}

\date{\today}

\begin{abstract}
This Article presents a nonequilibrium thermodynamic theory for the mean-field precipitation, aggregation and pattern formation of colloidal clusters. A variable gradient energy coefficient and the arrest of particle diffusion upon ``jamming" of cluster aggregates in the spinodal region predicts observable gel patterns that, at high inter-cluster attraction, form system-spanning, out-of-equilibrium networks with glass-like, quasi-static structural relaxation. For reactive systems, we incorporate the free energy landscape of stable pre-nucleation clusters into the Allen-Cahn-Reaction equation. We show that pattern formation is dominantly controlled by the Damk\"{o}hler number and the stability of the clusters, which modifies the auto-catalytic rate of precipitation. As clusters individually become more stable, bulk phase separation is suppressed.

\begin{description}
\item[Keywords] {\it nonequilibrium thermodynamics, cluster-cluster aggregation, reaction-diffusion, dynamic arrest, phase-field modeling}
\end{description}
\end{abstract}

\maketitle

\section{Introduction}
A colloid is a collection of nanometer- to micron-sized particles interacting in a fluid or solution. There has been much interest in studying colloids due to their ability to mimic atomic systems inaccessible to microscopy~\cite{lu2013colloidal}, and configure into functional, self-assembling structures~\cite{kraft2012surface, grzelczak2010directed, zhou2012self}. For instance, the colloidal nature of cement paste, a material of vast societal importance, has only recently been exploited to gain insight into the characteristics that lend it its exceptional mechanical properties~\cite{jennings2000model,masoero2012nanostructure,ioannidou2016mesoscale}. Likewise, the discharge products of lithium-ion batteries are being engineered to maximize ion transport and increase energy storage~\cite{girishkumar2010lithium,horstmann2013precipitation}, and magnetic nanoparticles are being functionalized as drug delivery vehicles, sealants, and separation aids~\cite{lattuada2007functionalization,ditsch2005controlled,baumgartner2013nucleation}. 

Recent experiments on the thermodynamics of reactive colloids have demonstrated pathways toward amorphous or crystalline bulk structures \textit{via} precipitation of stable prenucleation clusters, and reconciled these findings with classical nucleation theory~\cite{baumgartner2013nucleation}. In fact, two-step nucleation from stable precursors has been demonstrated in a host of particulate and biomineral systems~\cite{manoharan2003dense,carcouet2014nucleation,gebauer2008stable,tan2014visualizing,zhang2007does,privman1999mechanism}. Stabilizing mechanisms such as long-range electrostatic forces~\cite{zhang2012non, campbell2005dynamical, sciortino2004equilibrium,stradner2004equilibrium}, favorable ion coordination~\cite{gebauer2008stable}, and polar or micellar association~\cite{israelachvili2011intermolecular} allow persistent intermediates to form that settle into bulk structures upon super-saturation. Yet no physically consistent, systematic study has been brought forth to investigate the influences that control phase separation in these systems.

In this Article, we examine the mean-field non-equilibrium thermodynamics of reactive colloids that form mesoscale structures by aggregation and precipitation of stable precursors. Before introducing the reaction rate, it is shown that a convex gradient energy penalty reproduces characteristics akin to viscoelastic phase separation~\cite{tanaka1996universality, tanaka1997phase}, where contrasting entropic driving forces rather than differing constitutive behavior summon a rich set of gel-patterns also observable in nature. Dynamic asymmetry between the low-density gas and high-density gel phases is imposed by arresting particle diffusion at local percolation in the spinodal region, allowing a quasi-static system spanning gel to form. Next, we show that once reactive kinetics are included, the heterogeneity of the system is principally controlled by the Damk\"{o}hler number --- the ratio between the reaction rate and the cluster diffusion rate --- and the stability of the cluster intermediates. In particular, the thermodynamic landscape of clusters fully parameterizes a generalized Eyring reaction rate that enhances or suppresses bulk nucleation from solution~\cite{baumgartner2013nucleation, gebauer2014pre}.

\begin{figure*}
\begin{subfigure}[b]{0.32\textwidth}
\includegraphics[width=1.0\textwidth]{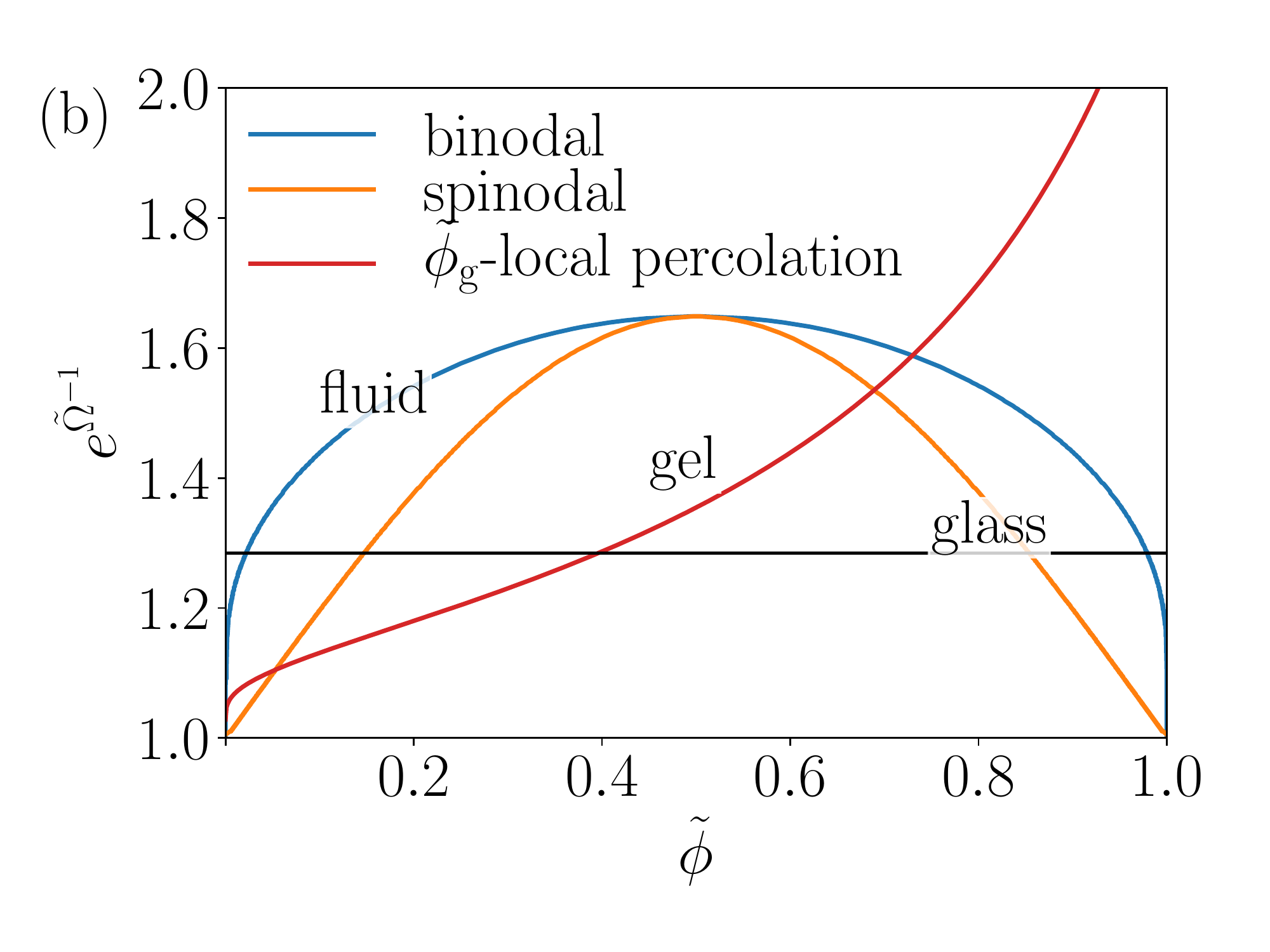}
\end{subfigure}
\begin{subfigure}[b]{0.32\textwidth}
\includegraphics[width=1.0\textwidth]{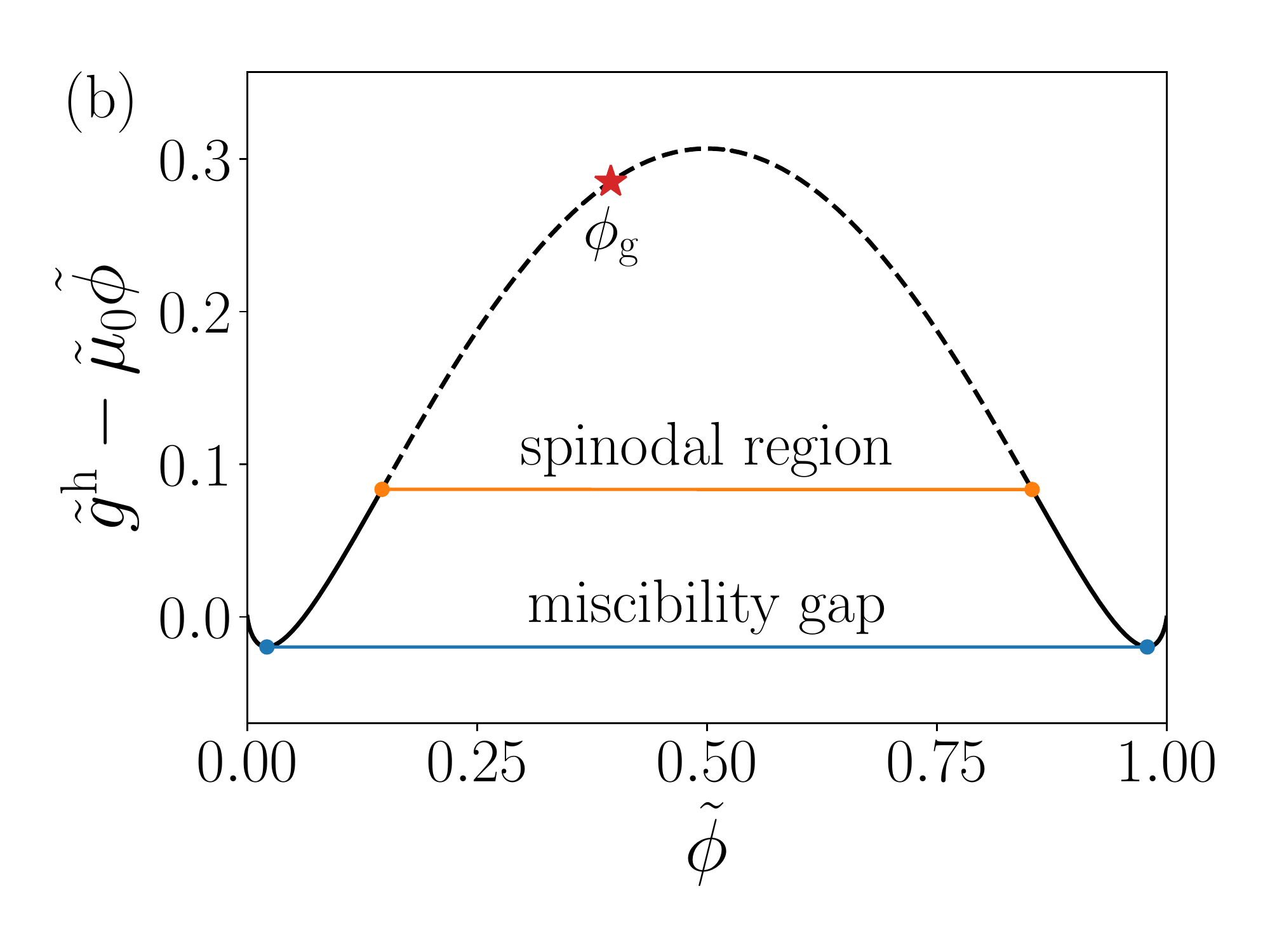}
\end{subfigure}
\begin{subfigure}[b]{0.32\textwidth}
\includegraphics[width=1.0\textwidth]{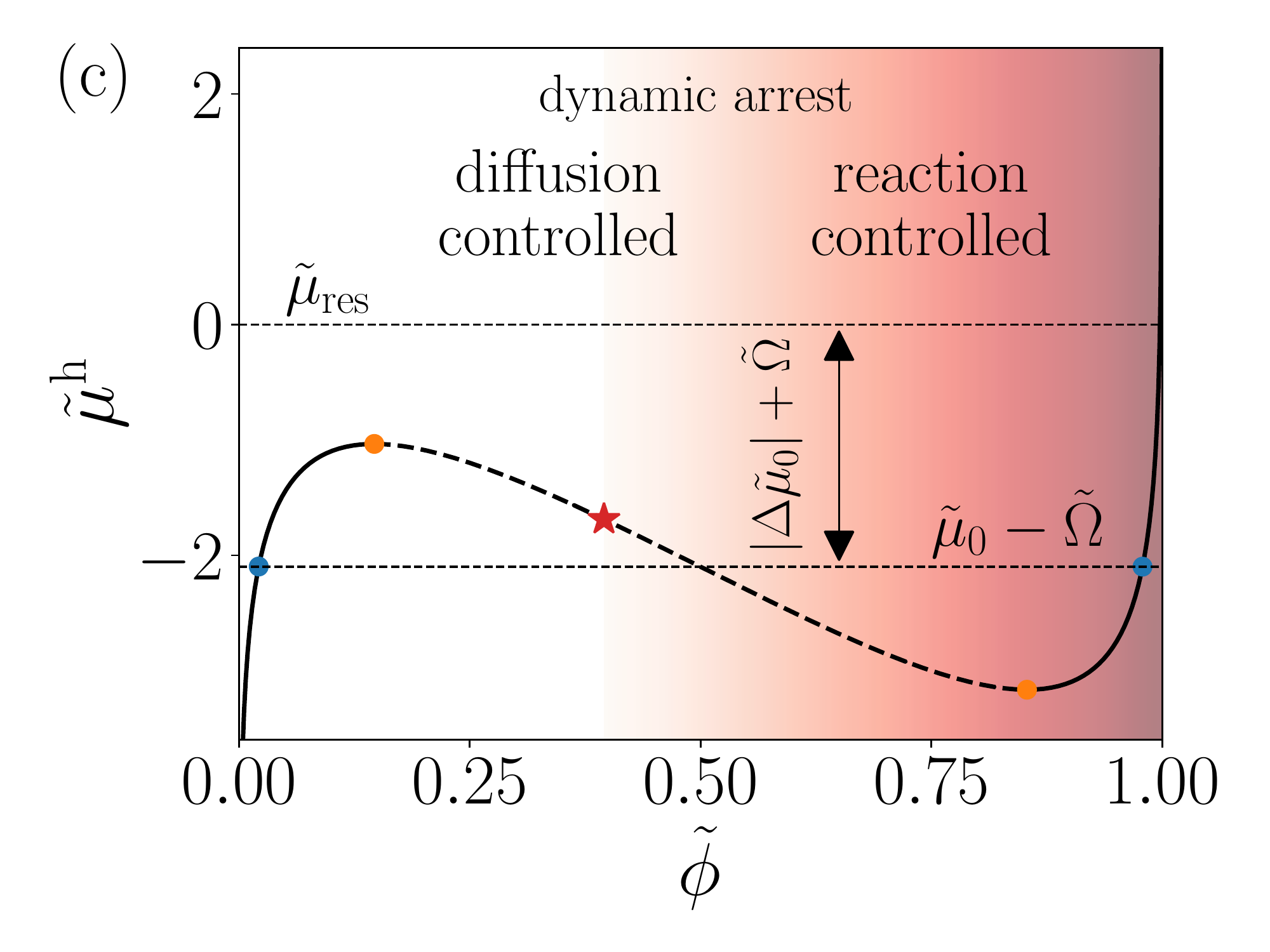}
\end{subfigure}
\caption{(a) Stability diagram depicting the binoal, spinodal, and gelation lines at varying quench depths; the gelation line corresponds to Eq.~\ref{eq:percolation}, and the black line indicates $\tilde{\Omega}=4.0$ selected for panels (b) and (c), and all simulation results of nonconserved systems. (b) Homogeneous part of Gibbs free energy of the clusters $\tilde{g}_\mathrm{h}$, where the linear term $(\tilde{\Omega}-\tilde{\mu}_0)\tilde{\phi}$ has been added for clarity. (c) Homogeneous part of the diffusional chemical potential $\tilde{\mu}_\mathrm{h}$, with $\tilde{\mu}_0 - \tilde{\Omega}$ and reservoir $\tilde{\mu}_\mathrm{res}$ indicated by dashed, horizontal lines. The red star indicates the packing fraction at the gelation transition.}
\label{fig:1}
\end{figure*}

\section{Nonequilibrium thermodynamics of attractive colloids}

Experimental observations and mode-coupling theory have demonstrated that two length scales dominate the physics of attractive colloids~\cite{bergenholtz1999nonergodicity, kroy2004cluster, cates2004theory}: Colloids aggregate into \textit{clusters} of characteristic size, which further assemble into an arrested mesocopic network. Upon quenching --- that is, rapidly increasing the relative attractive strength between particles, for instance by decreasing the temperature or changing the constitution of the solvent --- these systems undergo glass-like dynamic arrest where cluster-cluster aggregation exhibits limited bond-breakage and the structure factor $S_q$ does not significantly change on observational time scales. Akin to an athermal granular medium, the colloid-rich phase undergoes a jamming transition due to the local crowding of \textit{cluster aggregates}, as opposed to grains, which presents a ``hidden" binodal at densities far below the thermodynamically predicted dense equilibrium~\cite{cardinaux2007interplay, segre2001glasslike, gasser2001real}. As an example, this gelation line is drawn in red in the stability diagram shown in Fig.~\ref{fig:1}(a)). For mechanically unperturbed systems, thermal fluctuations promote densification toward equilibrium so slowly that further advances are made principally by reactive precipitation into the local structure. 

In particle systems with purely attractive interactions, ergodicity breaking at the scale of clusters creates polydisperse cluster sizes. However, if weak, long-range repulsive forces are introduced, simulations and experiments have shown colloids to form dilute suspensions of stable clusters with a narrow size distribution that persists over days~\cite{sciortino2004equilibrium, campbell2005dynamical}. Only after further increasing the mean packing density or attractive strength does further aggregation and phase separation proceed, leading to eventual arrest at the scale of cluster aggregates. To model the physics of these two length scales, we introduce a dynamic reaction-diffusion equation for colloidal clusters. Clusters are treated as renormalized particles of characteristic size $a$, whose local packing density is advanced by a field variable $\phi$ and diffusion is driven by gradients in the chemical potential of the clusters $\mu$. The ability of the dynamic equation to evolve realistic density patterns of colloidal gels is first demonstrated on conserved systems, where an initially homogeneous density field is assigned, and no additional insertion or deletion of clusters is permitted. These systems phase separate into high- and low-density regions, where clusters, initially disconnected from one another, form cluster aggregates, whose diffusivity is exponentially reduced at local percolation. Though we admit that polydispersity in sizes of cluster aggregates are relevant to the dynamics~\cite{zaccarelli2007colloidal,ioannidou2016crucial}, and that the dynamics are a history dependent function of $\phi$, the present study aims to reduce model complexity by focusing on the mesoscopic parameters that predict instability and pattern formation. Thus, we do not explicitly track the size distribution of the cluster aggregates; we reserve a forthcoming study to add an additional microscopic order parameter to investigate the elasticity of colloids in a continuum setting. We continue our study by deriving a general reaction rate to form clusters that are individually stable, but may collectively phase separate if a mesoscopic energy barrier is crossed. Specifically, the Allen-Cahn-Reaction equation, which was first introduced into electrochemistry to model phase separation in lithium-ion batteries~\cite{bazant2012theory, bai2011suppression}, utilizes the thermodynamic landscape of the stabilized clusters to measure the net rate of insertion. In other words, we model bulk nucleation as a two-step process. Throughout this article we refer to \textit{clusters} as stabilized base units that assemble into an out-of-equilibrium, mesoscopic gel network.

\subsection{Dynamic equation for density patterns}

We posit the internal chemical potential of a cluster $\mu(\phi,\boldsymbol{\nabla}\phi)$ as nonuniform, depending principally on the local packing density $\phi$ and its gradient $\boldsymbol{\nabla}\phi$~\cite{cahn1958free}. Thus, local cluster rearrangements are driven by spatial variations in $\mu$, and the evolution of the system is modeled by a general reaction-diffusion equation for nonequilibrium thermodynamic mixtures~\cite{bazant2012theory, cahn1958free, allen1979microscopic},
\begin{equation}
    \frac{\partial \tilde{\phi}}{\partial t} = \boldsymbol{\nabla} \cdot \left(\frac{D\tilde{\phi}}{k_\mathrm{B}T} \boldsymbol{\nabla} \mu\right) + R(\tilde{\phi},\mu,\mu_\mathrm{res}),
\label{eq:reac_diff}
\end{equation}
where $k_\mathrm{B}T$ sets the thermal energy scale, $D(\tilde{\phi})$ is the tracer diffusivity, $R$ is a reaction rate controlling insertion or deletion of clusters, and $0\le \tilde{\phi} = \phi/\phi_\mathrm{m}\le1$ is the filling fraction with $\phi_\mathrm{m}$ the maximum packing density ($\,\,\tilde{}\,$ is henceforth used to signify nondimensionalized and normalized quantities). The first, conserved term is a Cahn-Hilliard kernel that tracks cluster diffusion within the domain~\cite{cahn1958free}, and the second, nonconserved term is an Allen-Cahn reaction rate that acts as a cluster source or sink~\cite{allen1979microscopic,bazant2012theory}. While diffusion depends only on the local chemical potential, the reaction rate in this open system depends also on the external reservoir potential $\mu_\mathrm{res}$. The explicit expression for $R$ will be derived in a section below.

\begin{figure*}
\begin{subfigure}[b]{0.36\textwidth}
\includegraphics[width=1.0\textwidth]{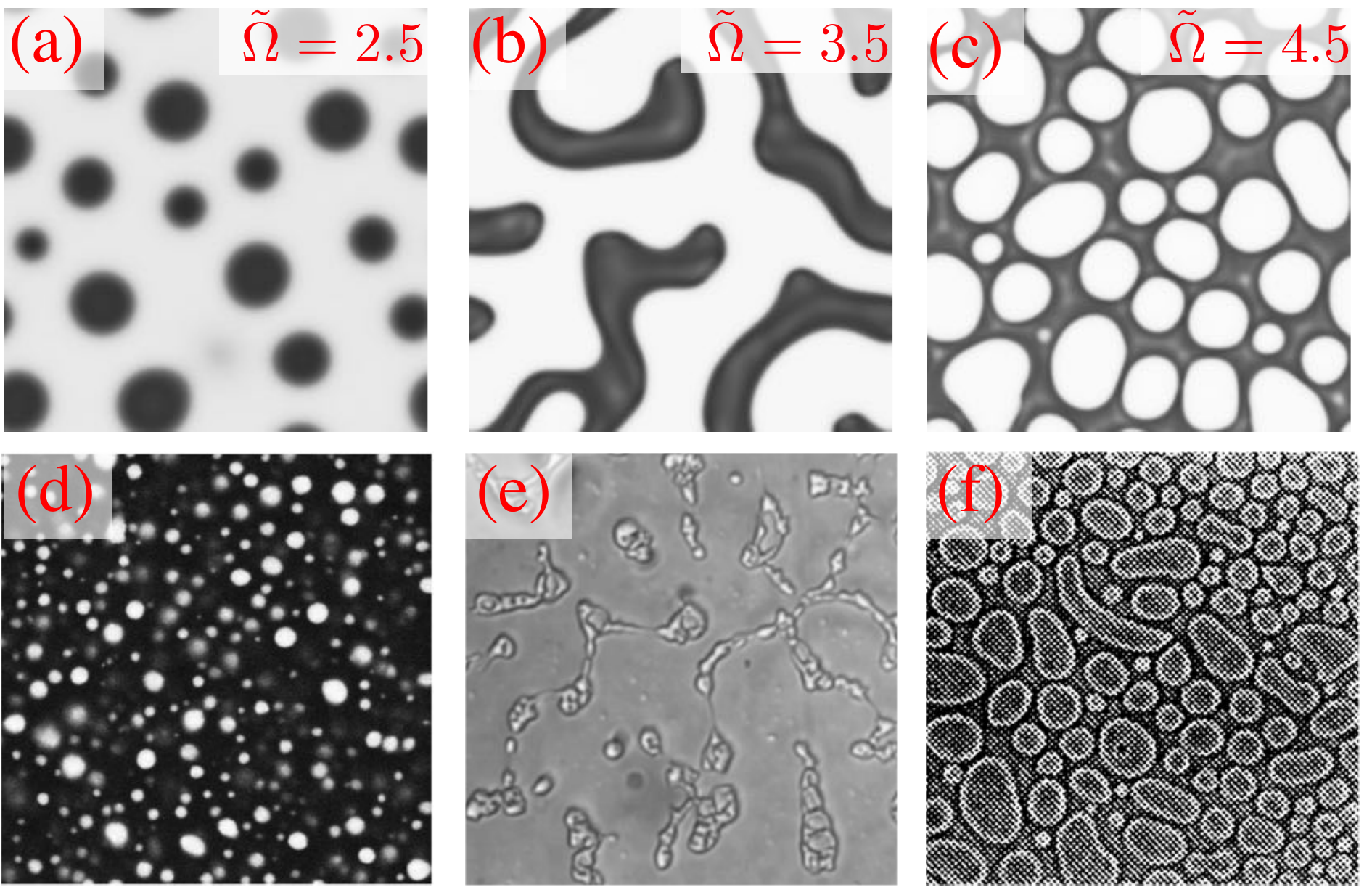}
\end{subfigure}
\begin{subfigure}[b]{0.30\textwidth}
\includegraphics[width=1.0\textwidth]{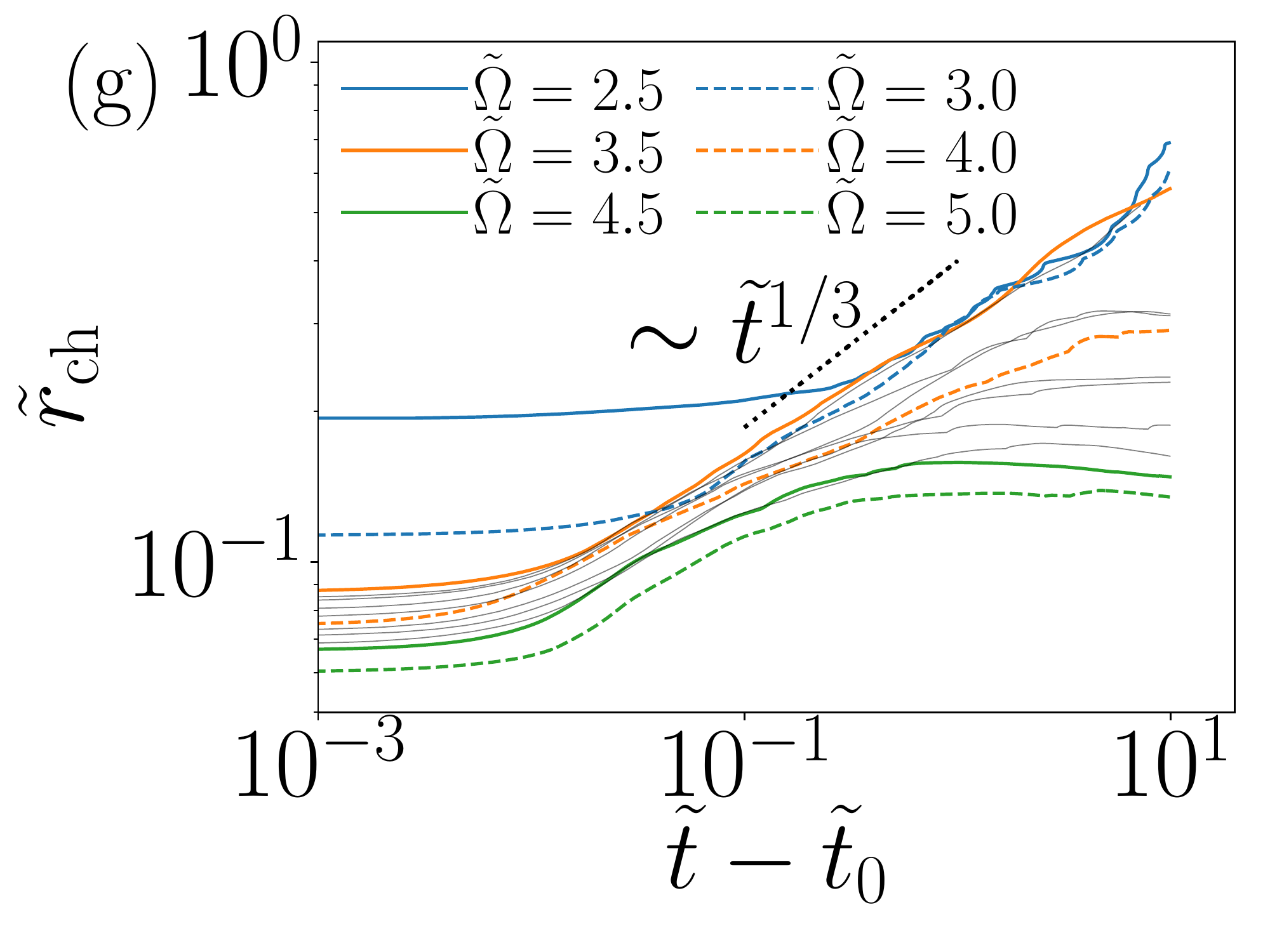}
\end{subfigure}
\begin{subfigure}[b]{0.32\textwidth}
\includegraphics[width=1.0\textwidth]{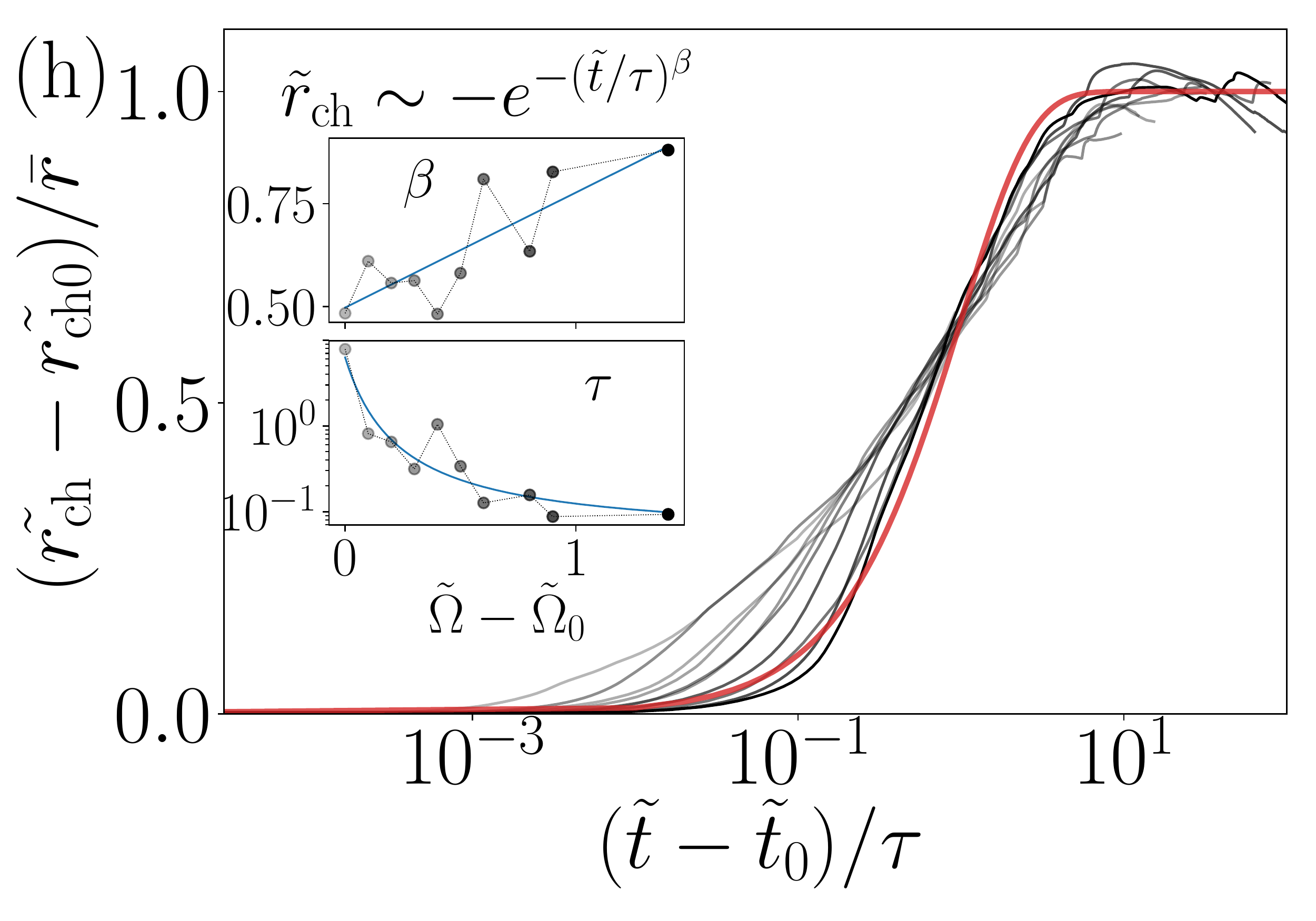}
\end{subfigure}
\caption{(a)-(c) Snapshots of $\tilde{\phi}(\mathbf{x})$ for conserved colloidal systems at $\tilde{t} = t L_\mathrm{sys}^2/D_0 = 1.0$ for varying quench depths $\tilde{\Omega}$; darkness is proportional to $\tilde{\phi}$. (d)-(e) Experimental images for (d) the demixing of milk protein~\cite{de2002phase}, (e) polystyrene-poly(vinyl methyl ether) \cite{el2007phase}, and (f) polystyrene-diethyl malonate solutions~\cite{tanaka1993unusual}. (g) Time evolution of the characteristic domain size $\tilde{r}_\mathrm{ch}$, where $\tilde{t}_0$ denotes the time of maximum interface area. (h) Evolution of the characteristic domain size rescaled to show its stretched exponential behavior; $\tilde{r}_\mathrm{ch}^0$ corresponds to the domain size at $\tilde{t}_0$, and the insets show $\beta \simeq C_0 (\Omega - \tilde{\Omega}_0)$ and $\ln(\tau) \simeq C_1(\Omega - \tilde{\Omega}_0) /(C_2 +(\Omega - \tilde{\Omega}_0))$ with fitting constants $C_i$. Simulations were run at numerical resolution $512 \times 512$ and system size $L_\mathrm{sys}=1$ with parameters $\kappa_0/n_\mathrm{s} k_\mathrm{B} T L_\mathrm{sys}^2=0.001$, $\chi = 3.0$, $\gamma = 1.2$, $\nu=0.88$, $\tilde{\xi}_\mathrm{g} = 50$, and $\phi^0_\mathrm{g}/\phi_\mathrm{m}=0.4$.}
\label{fig:2}
\end{figure*}

To calculate $\mu$, the free energy landscape of a system of volume $V$ is measured using a Ginzburg-Landau functional,
\begin{equation}
\label{eq:free_energy}
    G = \int_V g(\tilde{\phi},\nabla\tilde{\phi})\mathrm{d}V = \int_V \left(g^{\mathrm{h}}(\tilde{\phi}) + \frac{\kappa(\tilde{\phi})}{2}|\nabla \tilde{\phi}|^2\right)\mathrm{d}V
\end{equation}
where Gibbs energy density $g$ is expressed as a sum of \textit{homogeneous} and \textit{inhomogeneous} contributions. Because the energy demanded in separating monomers quenched into clusters far exceeds that needed to separate clusters themselves, and we aim to model stabilized clusters that form at a characteristic size, the \textit{homogeneous} free energy density is expressed as a regular solution of cluster occupied sites and vacancies
\begin{equation}
\frac{g^\mathrm{h}(\tilde{\phi})}{n_\mathrm{s}k_\mathrm{B}T}=  \tilde{\phi} \ln(\tilde{\phi}) + (1-\tilde{\phi}) \ln(1 - \tilde{\phi}) - \tilde{\Omega} \tilde{\phi}^2 + \tilde{\phi} \tilde{\mu}_0
\label{eq:free_energy_density}
\end{equation}
with site density $n_\mathrm{s}$~\cite{bazant2012theory}. Thus, clusters act as renormalized particles, where $\tilde{\Omega}=\Omega/k_\mathrm{B}T$ and $\tilde{\mu_0}=\mu_0/k_\mathrm{B}T$ are, respectively, the interaction parameter between clusters and chemical potential of a cluster in a dilute system, each measured with respect to the thermal energy scale. The double-well shape of $g^\mathrm{h}$ is plotted in Fig.~\ref{fig:1}(b), where two minima define low- and high-density thermodynamic equilibria, and the concave spinodal region indicates packing densities at which homogeneous base states in conserved systems are unstable to density fluctuations. The variable gradient energy coefficient for the \textit{inhomogeneous} free energy contribution is chosen as 
\begin{equation}
\frac{\kappa(\tilde{\phi})}{n_\mathrm{s}a^2k_\mathrm{B}T} = \frac{2}{9}\frac{\tilde{\kappa}_0}{\phi(1-\phi)},
\label{eq:gradient_penalty}
\end{equation}
and encodes the interface width $l\simeq (2/3)\sqrt{\kappa_0/\Omega}$ and surface energy of the density field. De Gennes derived the expression in Eq.(\ref{eq:gradient_penalty}) by relating the response of fluctuations in the entropic portion of a free energy density similar to Eq.(\ref{eq:free_energy_density}) to the static structure factor and radius of gyration of polymer coils~\cite{de1980dynamics,de1979scaling}. Here, we repurpose his result for our attractive colloids, where erdogicity breaking at the scale of the clusters sets the length scale of the density fluctuations. With Gibbs free energy defined, the \textit{diffusional} chemical potential is calculated using the Euler-Lagrange equation~\cite{bazant2012theory}: 
\begin{widetext}
    \begin{equation}
            \frac{\mu}{k_\mathrm{B}T} =\frac{\partial \tilde{g}}{\partial \phi} - \boldsymbol{\nabla}\cdot \frac{\partial \tilde{g}}{\partial \boldsymbol{\nabla}\tilde{\phi}} = \ln\left(\frac{\tilde{\phi}}{1-\tilde{\phi}}\right) - 2\tilde{\Omega}\tilde{\phi} + \tilde{\mu}_0 + \frac{\tilde{\kappa}'(\tilde{\phi})}{2}|\tilde{\boldsymbol{\nabla}}\tilde{\phi}|^{2}-\tilde{\boldsymbol{\nabla}}\cdot\left(\tilde{\kappa}(\tilde{\phi})\tilde{\boldsymbol{\nabla}}\tilde{\phi}\right).
        \label{eq:chemical_potential}
    \end{equation}
\end{widetext}
This is the continuum analog to the standard definition of the chemical potential in particle-resolved systems. Above, $'$ denotes ordinary differentiation, and the gradient operator $\tilde{\nabla}=L_\mathrm{sys}\nabla $ has been normalized by the linear size of the system $L_\mathrm{sys}$. The shape of the homogeneous part of the chemical potential $\tilde{\mu}^\mathrm{h}=\mathrm{d}\tilde{g}^\mathrm{h}/\mathrm{d}\tilde{\phi}$ is plotted in Fig.~\ref{fig:1}(c), and will be discussed in relation to the reactive system's stability in greater depth below.

\subsection{Glass-like arrest of particle diffusion}

Several experimental investigations of phase separating colloids have shown that dynamic arrest is initiated when cluster aggregates crowd their local volume and the rate of bond formation exceeds the rate of bond breakage at a gelation line aptly described by~\cite{segre2001glasslike, gibaud2011phase}.

\begin{equation}
    \phi_\mathrm{g} = \phi_\mathrm{g}^0 \exp\left(-\frac{\Omega}{\chi k_\mathrm{B}T}\right).
    \label{eq:percolation}
\end{equation}
Above, $\chi$ is a constant of order unity, and $\phi_\mathrm{g}^0$ is a packing density near the glass transition of hard-spheres, here chosen as $\phi_\mathrm{g}^0 = 1.5\, \phi_\mathrm{m}$. As stated in Ref.\cite{segre2001glasslike}, the exponential form of Eq.(\ref{eq:percolation}) is suggestive of a thermally activated process, whereby $\phi_\mathrm{g}$ acts as the local packing at which the number of kinetic pathways toward a lower energy state drastically reduces. Approaching this threshold leads to a rapid increase of the mean size of the cluster aggregates and a sharp decrease in local particle diffusion. A plethora of details are relevant to accurately modeling particle motion, including the size distribution of the aggregates, its hydrodynamic interactions with the solvent~\cite{varga2016hydrodynamic}, the history dependence of the aggregation process, and dynamical heterogeneity that varies across orders of magnitude~\cite{zaccarelli2007colloidal}. Because this article investigates aggregation that is largely unidirectional, favoring net densification, and we wish to maintain model parsimony, we opt to introduce an expression for the tracer diffusivity that captures the general physics of motion, though admittedly needs to be parameterized for the system at hand. 

\begin{figure}
\begin{subfigure}[b]{0.4\textwidth}
\includegraphics[width=1\textwidth]{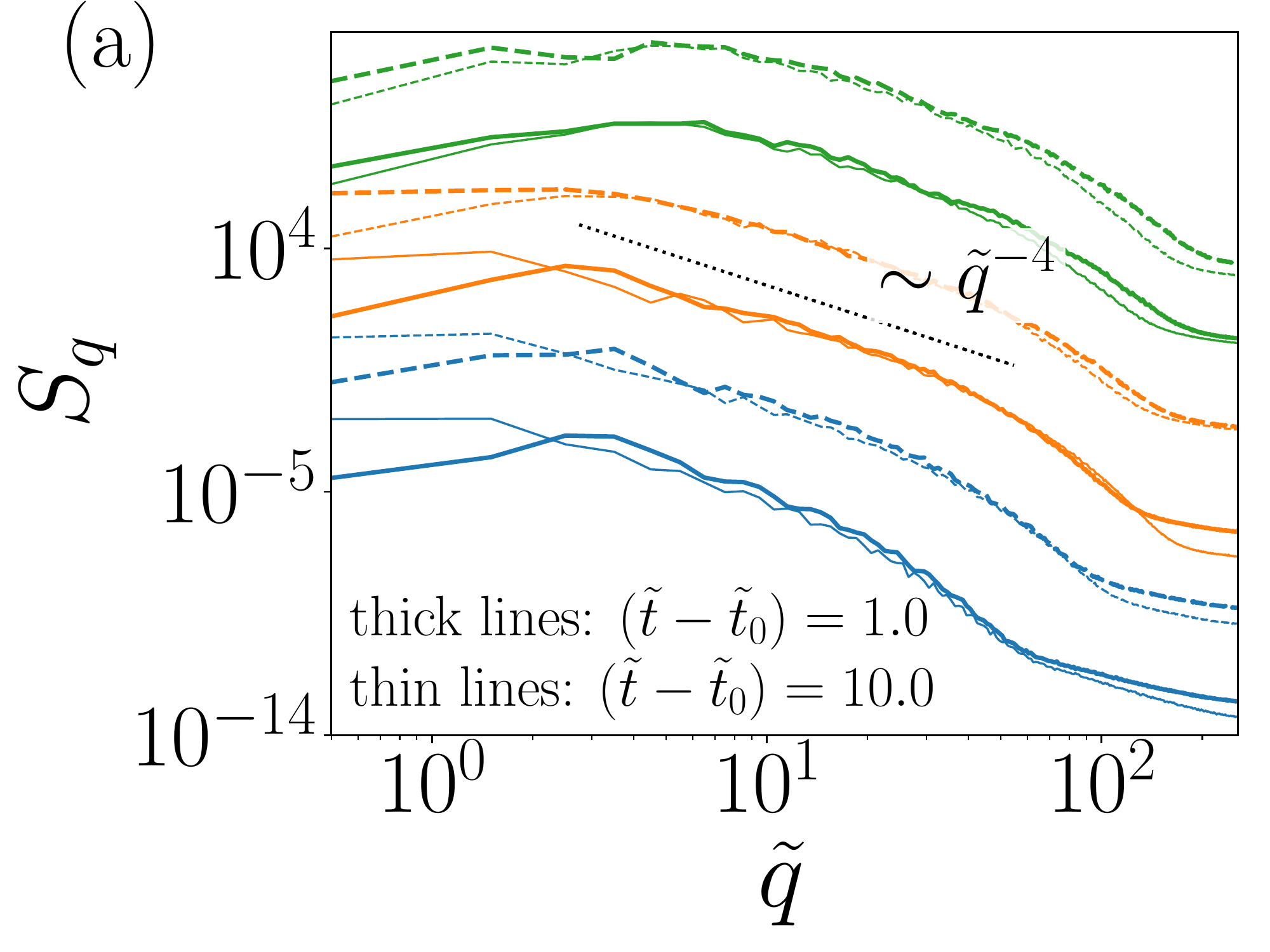}
\end{subfigure}
\begin{subfigure}[b]{0.4\textwidth}
\includegraphics[width=1\textwidth]{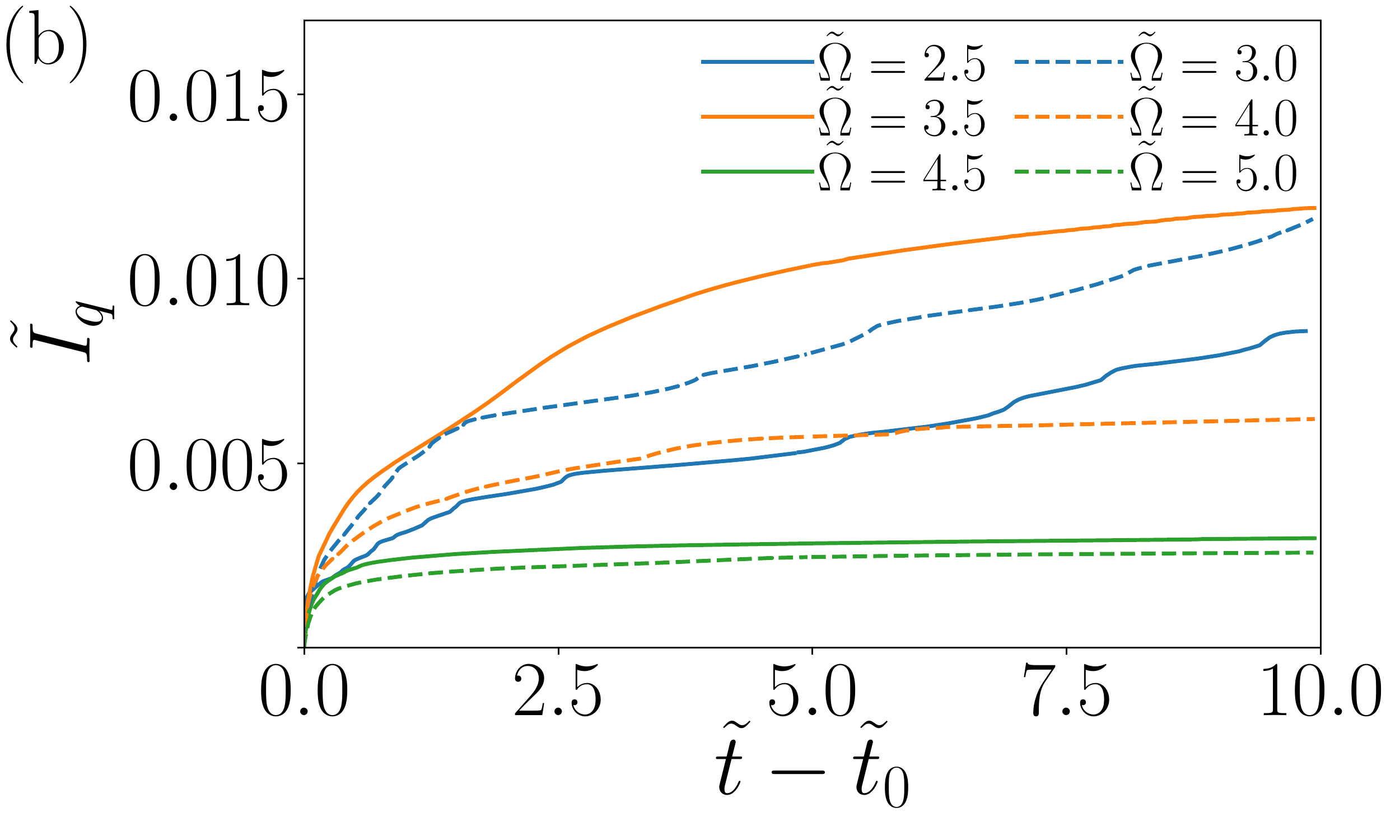}
\end{subfigure}
\caption{(a) Structure factor of $\tilde{\phi}(\mathbf{x})$ for various $\tilde{\Omega}$ at $\tilde{t}-\tilde{t}_0 = 1.0$ (thick lines) and $\tilde{t}-\tilde{t}_0 = 10.0$ (thin lines), where $\tilde{t}_0$ denotes the time at which the interface area is at its maximum; curves are vertically offset for clarity. (b) Time evolution of the peak integrated intensity $\tilde{I}_q = \int S_q \mathrm{d}\tilde{q}$; the legend indicating $\tilde{\Omega}$ also corresponds to the curves in (a). Dissolution and redoposition of small ``droplets" onto adjacent larger structures --- Ostwald ripening --- is marked by step changes in $I_q$ for $\tilde{\Omega}=2.5$ and $\tilde{\Omega}=3.0$. Time and length scales are normalized by the dilute limit self-diffusivity of a cluster $D_0$ and the system size $L_\mathrm{sys}$ as follows: $\tilde{t} = t D_0/ L_\mathrm{sys}^2$, and $\tilde{q} = q L_\mathrm{sys}$.}
\label{fig:3s}
\end{figure}

\begin{figure*}
\begin{subfigure}[b]{0.29\textwidth}
\includegraphics[width=1.0\textwidth]{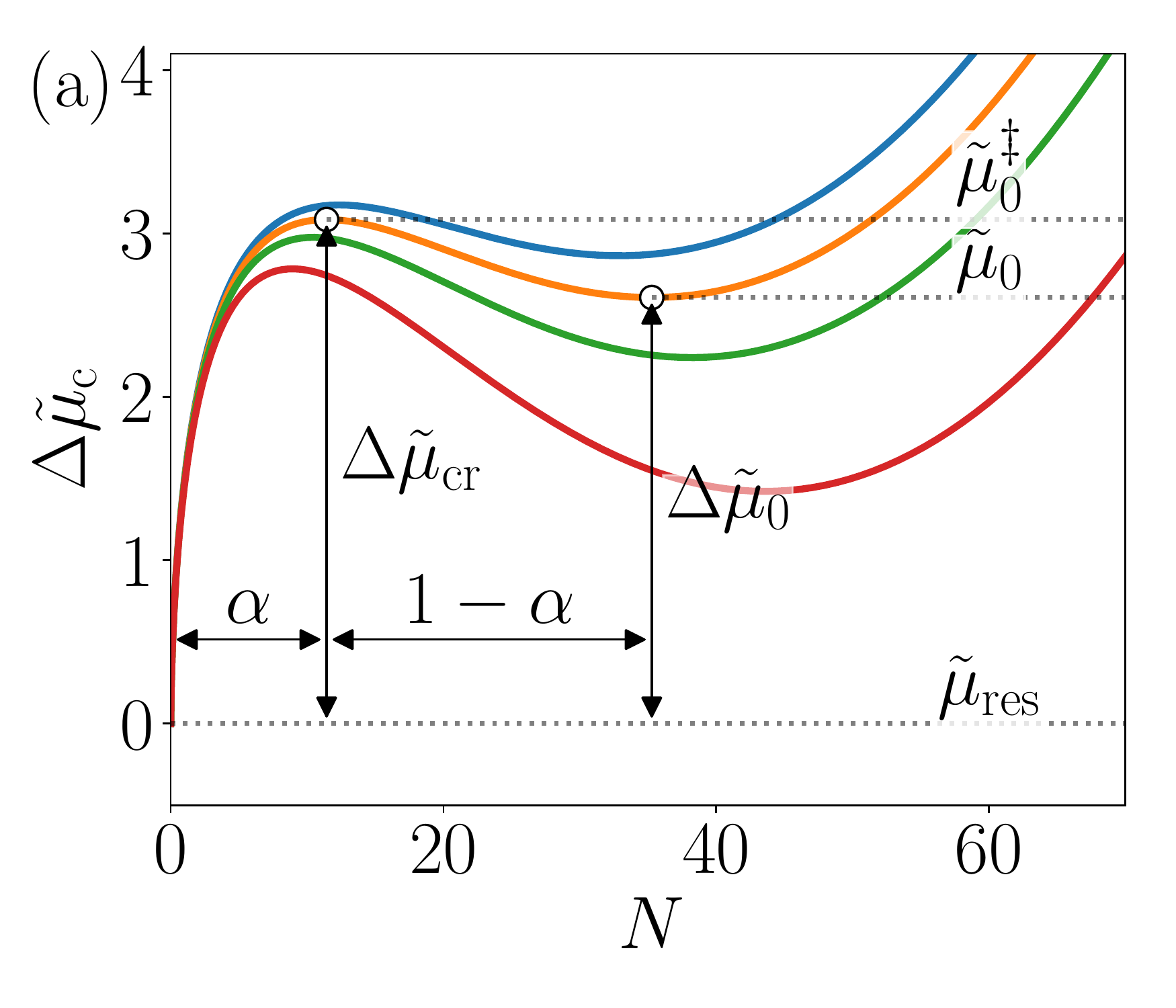}
\end{subfigure}
\begin{subfigure}[b]{0.34\textwidth}
\includegraphics[width=1.0\textwidth]{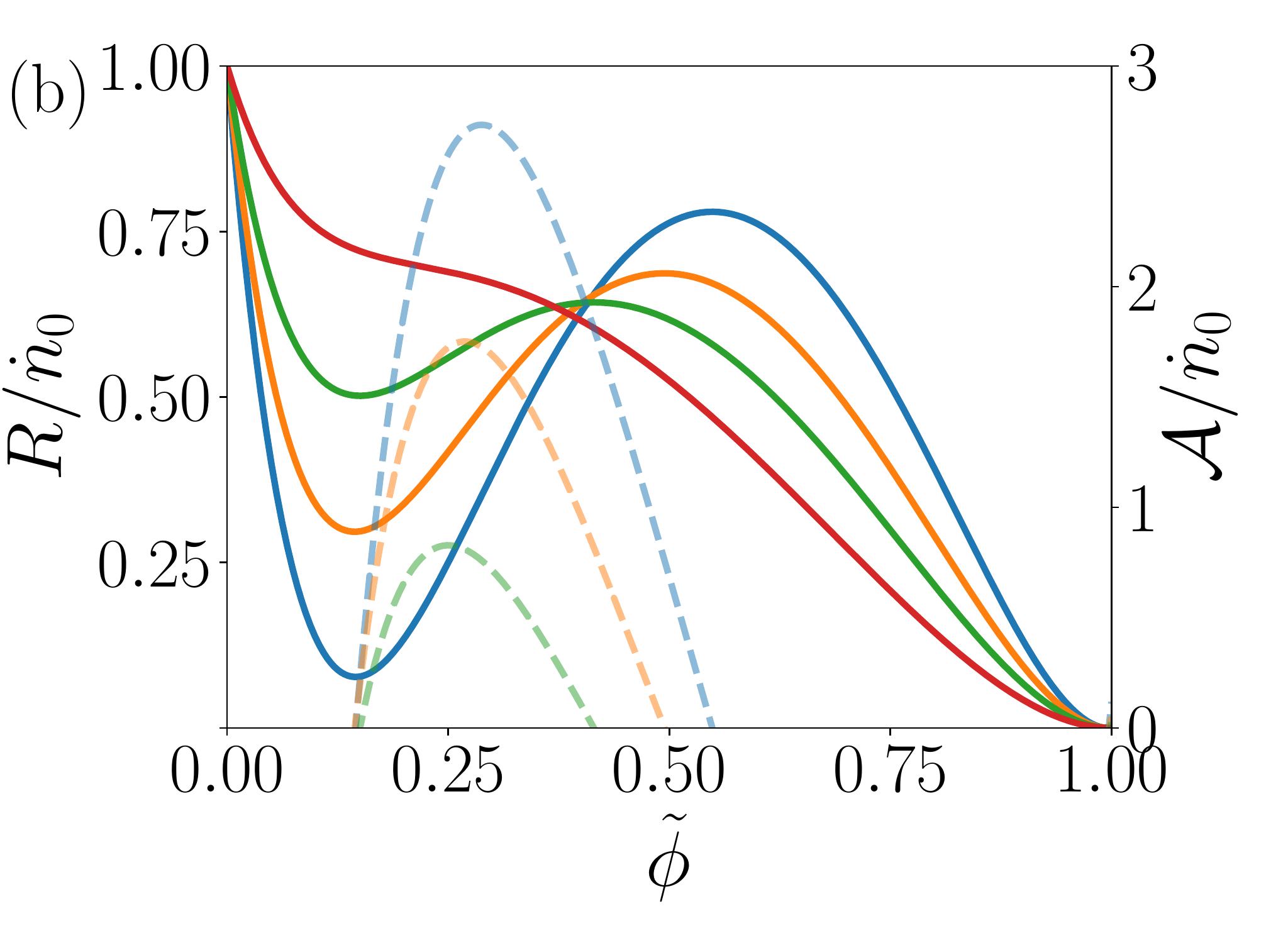}
\end{subfigure}
\begin{subfigure}[b]{0.29\textwidth}
\includegraphics[width=1.0\textwidth]{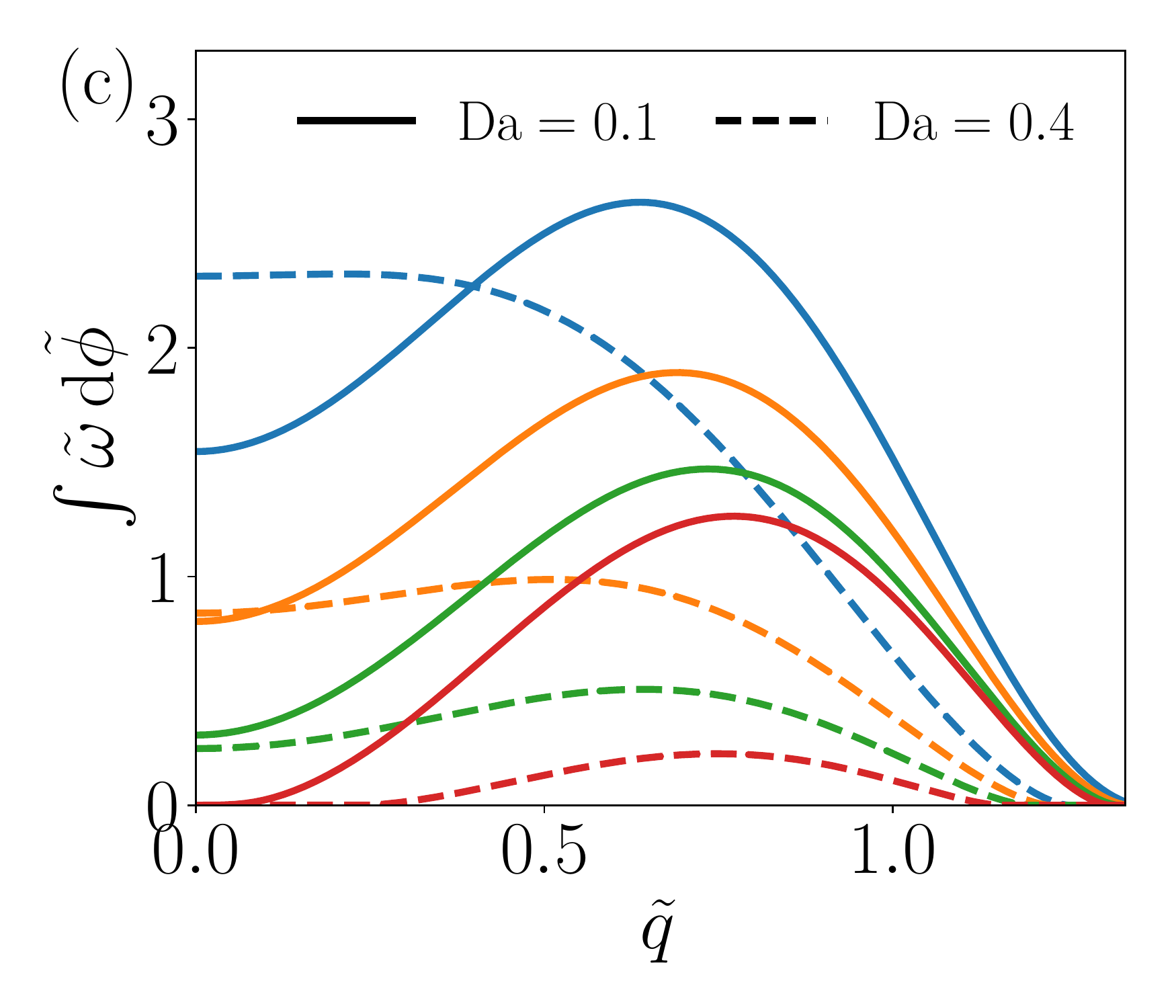}
\end{subfigure}
\caption{(a) Thermodynamic landscape of the transition state to form stable cluster precursors for varying $\Delta \mu_0^\mathrm{m}$ (see Eq.~\ref{eq:delta_mu}). (b) Reaction rate (left axis) and auto-catalytic rate (right axis) in function $\tilde{\phi}$. (c) Integrated growth rate of the $\tilde{q}$ Fourier mode as calculated from the $l^2$-norm of a perturbation for differing $\mathrm{Da}$. All plots correspond to $\tilde{\Omega}=4.0$ and line colors correspond to trials listed in Table~\ref{tab:params}: blue - T1, orange - T2, green - T3, and red - T4.}
\label{fig:3}
\end{figure*}

On approach of the gelation line from low densities, bulk measurements on attractive colloids show power-law divergence in viscosity~\cite{segre2001glasslike}, observations expected to be the result of particle motion that is increasingly collective~\cite{witten1981diffusion, wiltzius1987hydrodynamic, hess1986hydrodynamic}. As a consequence, we assume the diffusivity to scale as $D\sim(\xi/a)^{\gamma/\nu}\sim|\phi-\phi_\mathrm{g}|^{-\gamma}$ for $\phi < \phi_\mathrm{g}$, where $\gamma$ and $\nu$ are critical exponents for the diffusivity and correlation length $\xi$, respectively. Though diffusion drastically reduces once the percolation cluster forms, continued motion in physical systems persists due to the finite size of the system~\cite{stauffer1979scaling}, cluster aggregates that remain disconnected from the percolation cluster~\cite{meakin1983formation}, and activated events caused by internal stresses and thermal fluctuations~\cite{cipelletti2000universal}. Hence, we adjust the power-law relation for $\xi$ to its characteristic length at gelation, and infer $\phi = \phi_\mathrm{g}$ as the packing density at which the cluster aggregates percolate their cage of size $\xi_\mathrm{g}$. Specifically, we relate $\tilde{\phi}_\mathrm{g}$ and $\tilde{\xi}_\mathrm{g}=(\xi_\mathrm{g}/a)$ in our expression for the tracer diffusivity as follows:
\begin{equation}
\label{eq:diffusion_coefficient}
    D =  D_0 \left( \epsilon H(\epsilon) + \tilde{\xi}_\mathrm{g}^{-1/\nu}\exp\left(-\frac{\tilde{\xi}_\mathrm{g}^{1/\nu}}{2}\left|\epsilon\right|\right)\right)^{\gamma}.
\end{equation}
Above, $D_0\approx k_\mathrm{B}T/3\pi a \eta$ is the tracer diffusivity in a dilute system with $\eta$ being the solvent viscosity, $\epsilon = (\tilde{\phi}_\mathrm{g}-\tilde{\phi})/\tilde{\phi}_\mathrm{g}$ is the reduced density, and $H$ is the Heaviside function. The form of Eq.(\ref{eq:diffusion_coefficient}) imparts the following characteristics on $D$: i) a power-law dependence on the packing density at low $\tilde{\phi}$, ii) $\xi_\mathrm{g}$ as the relevant length scale at $\phi_\mathrm{g}$, and iii) a stretched exponential relaxation upon arrest~\cite{cipelletti2000universal}. Similar to Vogel-Fulcher-Tammann relaxation in glasses~\cite{adam1965temperature}, Eq.(\ref{eq:percolation}) ascribes dynamic arrest to the loss of thermally activated rearrangement, and Eq.(\ref{eq:diffusion_coefficient}) imposes an exponential decay in the collective diffusion of cluster aggregates upon jamming.

\subsection{Results for conserved systems}

Before deriving the reaction rate $R$, we exemplify the flexibility of our model in resolving pattern formation of conserved fields. Simulations were run by implementing the weak form of Eq.(\ref{eq:reac_diff}) in the open source Multiphysics Object-Oriented Simulation Environment (MOOSE)~\cite{MOOSE}, a finite element analysis software. Trials were run by initializing the density field at a homogeneous base state with mean filling fraction $\tilde{\Phi} = (\int_V \tilde{\phi}\, \mathrm{d}V)/V = 1/3$ and adding Langevin noise with variance scaled by $D$ to account for thermal fluctuations~\cite{hohenberg1977theory}. Figs.~\ref{fig:2}(a)-(f) compare snapshots of simulated and observed colloidal mixtures quenched into varying states of inter-cluster attraction. As displayed, adjusting inter-cluster attractive strength, $\tilde{\Omega}$, predicts dramatic changes to the texture and dynamics of the resulting fields despite fixing the mean filling fraction. The fields predict droplet nucleation at low $\tilde{\Omega}$, metastable filaments at moderate $\tilde{\Omega}$, and a spanning honeycomb network at high $\tilde{\Omega}$ (Figs.~\ref{fig:2}(a)-(c)), similarly seen in experimental analogs (Figs.~\ref{fig:2}(d)-(e)). For the relevant system, our model reproduces several features also observed in viscoelastic phase separation~\cite{tanaka1997phase,tanaka1996universality}: i) Nucleation of a colloid rich phase, ii) volume shrinking of the colloid rich phase and nucleation of holes therein, and iii) the formation of a spanning network. However, instead of attributing the kinetic asymmetry to differences in elasticity, here the asymmetry results from differences in the entropic penalty from density fluctuations. The convex shape of $\kappa$ and arrest of the colloid rich phase in the spinodal region dramatically reduces its interfacial tension, whence the field in Fig.~\ref{fig:2}(b) evolved into its quasi-stable nonspherical shape. For arrest at lower $\phi_\mathrm{g}$, the aggregates become increasingly cohesive, invading the solvent to form long-range bridges as is seen in Fig.~\ref{fig:2}(c). Unlike viscoelastic phase separation, spanning systems of colloidal aggregates become quiescent, showing little structural evolution over decades~\cite{lu2008gelation,varga2016hydrodynamic}, and phase inversion is not observed; simulations for $\tilde{\Omega}=\{4.5,5.0\}$ were run for a decade beyond the results shown in Fig.\ref{fig:2} without observing phase inversion.

For an isotropic density field, the structure factor is calculated as \begin{equation}
S_q(\tilde{q}) = \frac{\langle \hat{\phi}(\tilde{\mathbf{q}})\hat{\phi}^*(\tilde{\mathbf{q}}) \rangle}{\tilde{\Phi}},
\end{equation} where $\hat{\phi}(\tilde{q})$ is the Fourier transform of $\tilde{\phi}(\tilde{\mathbf{x}})$, $^*$ indicates complex conjugation, and brackets denote spherical averaging. Increased quench depths shift the peak frequency in $S_q$ toward shorter wavelengths, and a decline, rather than growth, of long wavelengths (low $\tilde{q}$) signifies a system-spanning gel of increasingly rigidifying, thinning ligaments as discerned in Fig.~\ref{fig:3s}(a). This finding is verified in Fig.~\ref{fig:2}(g), which depicts the evolution of the characteristic domain size measured from the first moment in $\tilde{q}$ as 
\begin{equation}
\tilde{r}_\mathrm{ch}=\left(\frac{\int\tilde{q} S_q\mathrm{d}\tilde{q}}{I_q}\right)^{-1},
\end{equation}
where $I_q = \int S_q \mathrm{d}\tilde{q}$ is the integrated peak intensity. For low $\tilde{\Omega}$, coarsening proceeds via Ostwald ripening (e.g., $\tilde{\Omega}=2.5$; Fig.~\ref{fig:2}(a)) and contraction of gel filaments (e.g., $\tilde{\Omega}=3.5$; Fig.~\ref{fig:2}(b)), adhering to early-stage diffusive $\tilde{t}^{1/3}$ power-law scaling for $\tilde{r}_\mathrm{ch}$~\cite{lifshitz1961kinetics,siggia1979late,bailey2007spinodal}. In fact, loss of step-changes in the evolution of $I_q$ in Fig.~\ref{fig:3s}(b) indicate a clear distinction in coarsening dynamics. As $\tilde{\Omega}$ is increased, diffusive scaling in $\tilde{r}_\mathrm{ch}$ is abandoned and domain growth arrests into an out-of-equilibrium gel with anticipated bulk elasticity~\cite{testard2011influence,grant1993volume}; here, percolation of the bulk volume is a prerequisite for arrest. Denoting $\tilde{\Omega}_0$ as the minimal quench depth to form a system spanning gel, Fig.\ref{fig:2}(h) demonstrates stretched exponential relaxation in the evolution of $r_\mathrm{ch}$ --- typical of glass forming colloids --- , where the Vogel-Fulcher-Tammann time scale $\tau$ and stretching exponent $\beta$ are functions of the deviation from the mesoscopic percolation threshold $\tilde{\Omega}-\tilde{\Omega}_0$~\cite{williams1955temperature}. As low stretching exponents $\beta < 1$ are the result of spatially heterogenous dynamics, and $\beta$ increases with $\tilde{\Omega}$, heterogeneous dynamics are most prominent in weakly attractive systems close to the gelation threshold.

\begin{table}
\caption{Parameters defining the free energy landscapes of the stable cluster precursors, where $\Delta \tilde{\mu}_\mathrm{c} = \Delta \tilde{\mu}_0^\mathrm{m} N +  B N^{2/3} +  C N^{5/3}$, and $B=0.023$ and $C = 2.608$ are left constant.}
\begin{ruledtabular}
\begin{tabular}{l | c c c}
\textbf{Trial}&
$\Delta \tilde{\mu}_0^\mathrm{m}$&
$\Delta \tilde{\mu}_0$&
$\alpha$
\\[0.1cm]
\textbf{T1}&
$-0.883$&
$2.864$&
$0.376$\\[0.1cm]
\textbf{T2}&
$-0.890$&
$2.608$&
$0.324$\\[0.1cm]
\textbf{T3}&
$-0.900$&
$2.240$&
$0.273$\\[0.1cm]
\textbf{T4}&
$-0.920$&
$1.422$&
$0.205$\\[0.1cm]
\end{tabular}
\end{ruledtabular}
\label{tab:params}
\end{table}

\subsection{Reaction rate governed by stable cluster precursors}

\begin{figure*}
\begin{subfigure}[b]{0.325\textwidth}
\includegraphics[width=1.0\textwidth]{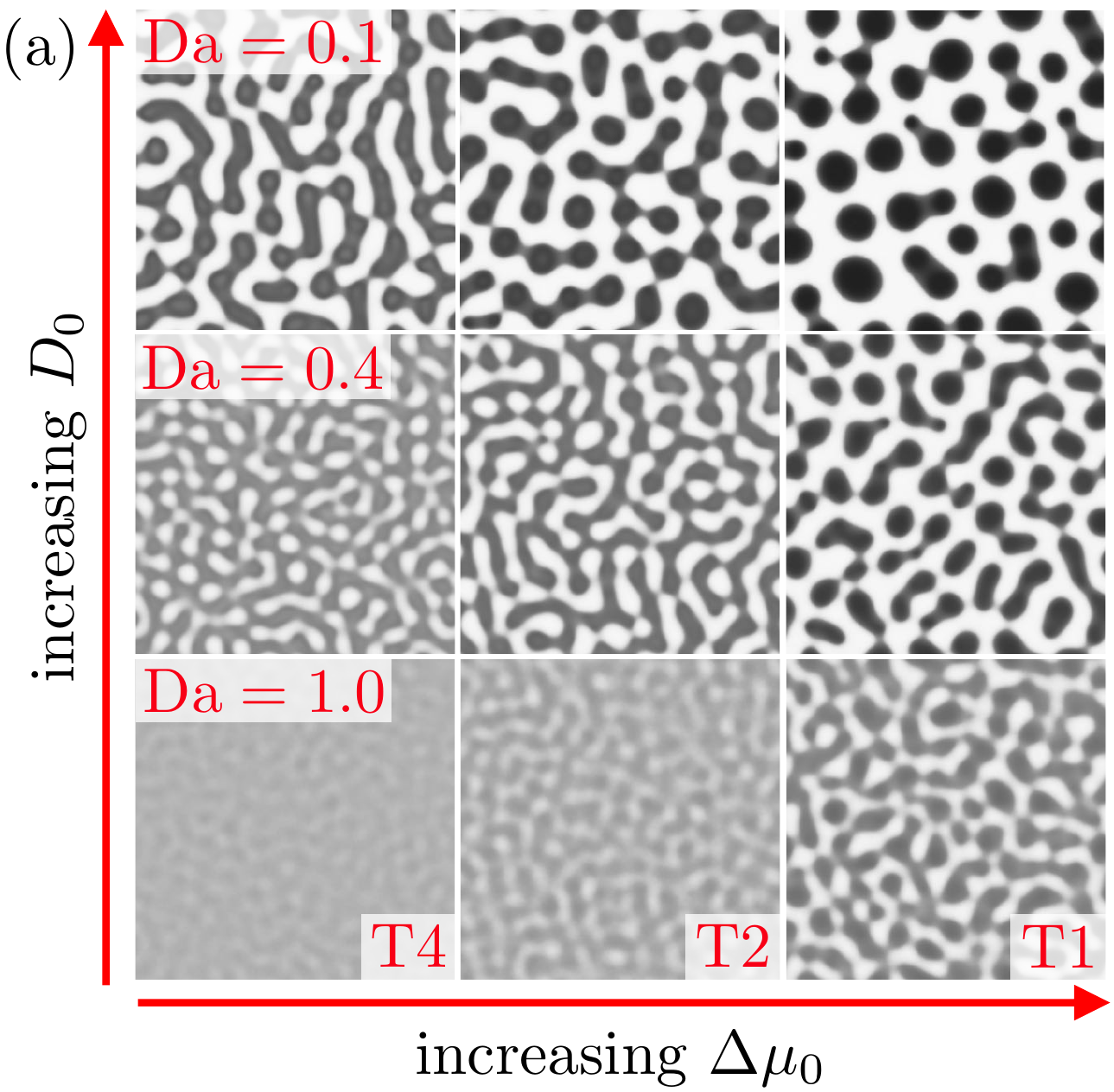}
\end{subfigure}
\begin{subfigure}[b]{0.325\textwidth}
\includegraphics[width=1.0\textwidth]{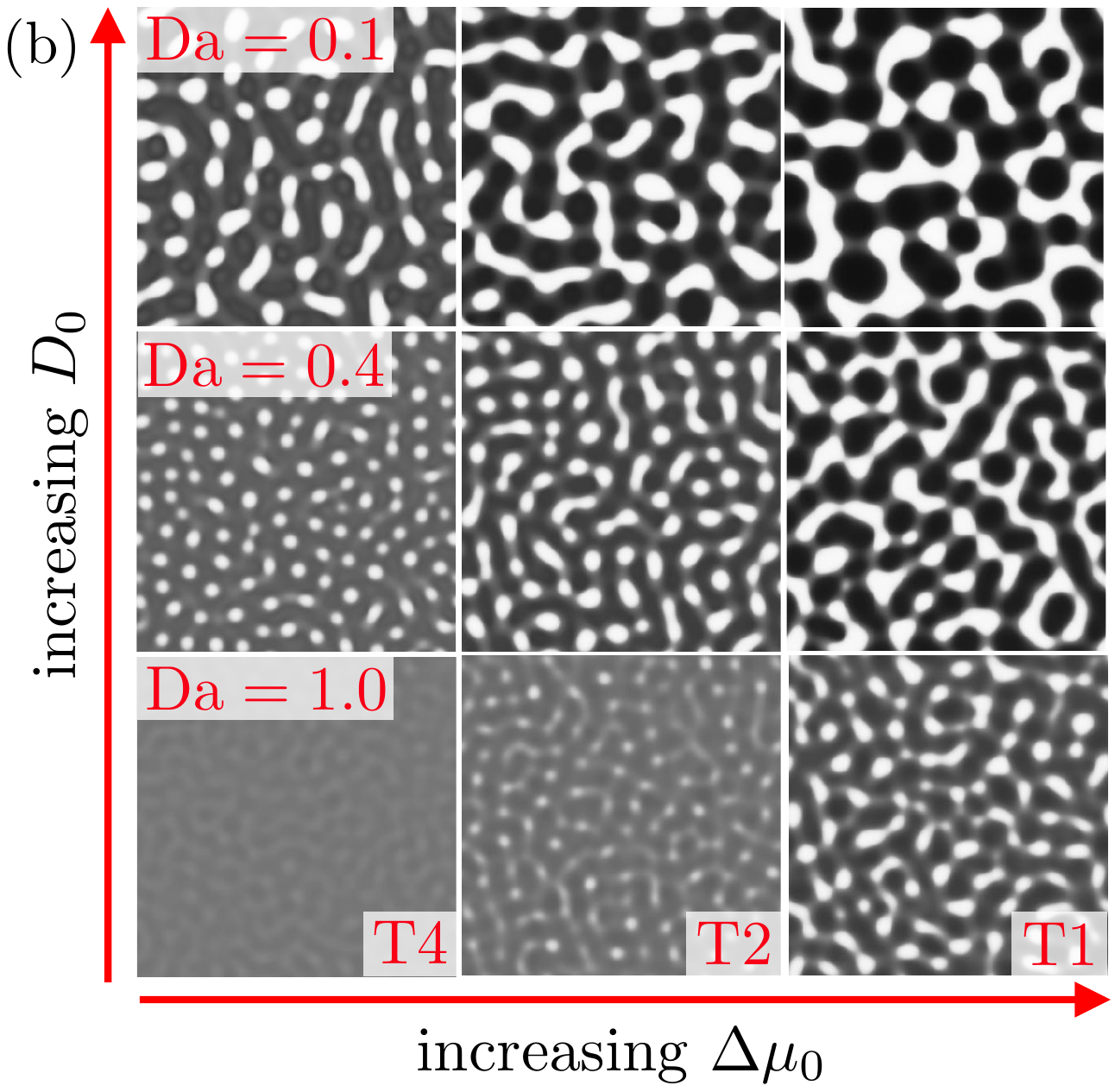}
\end{subfigure}
\begin{subfigure}[b]{0.325\textwidth}
\includegraphics[width=1.0\textwidth]{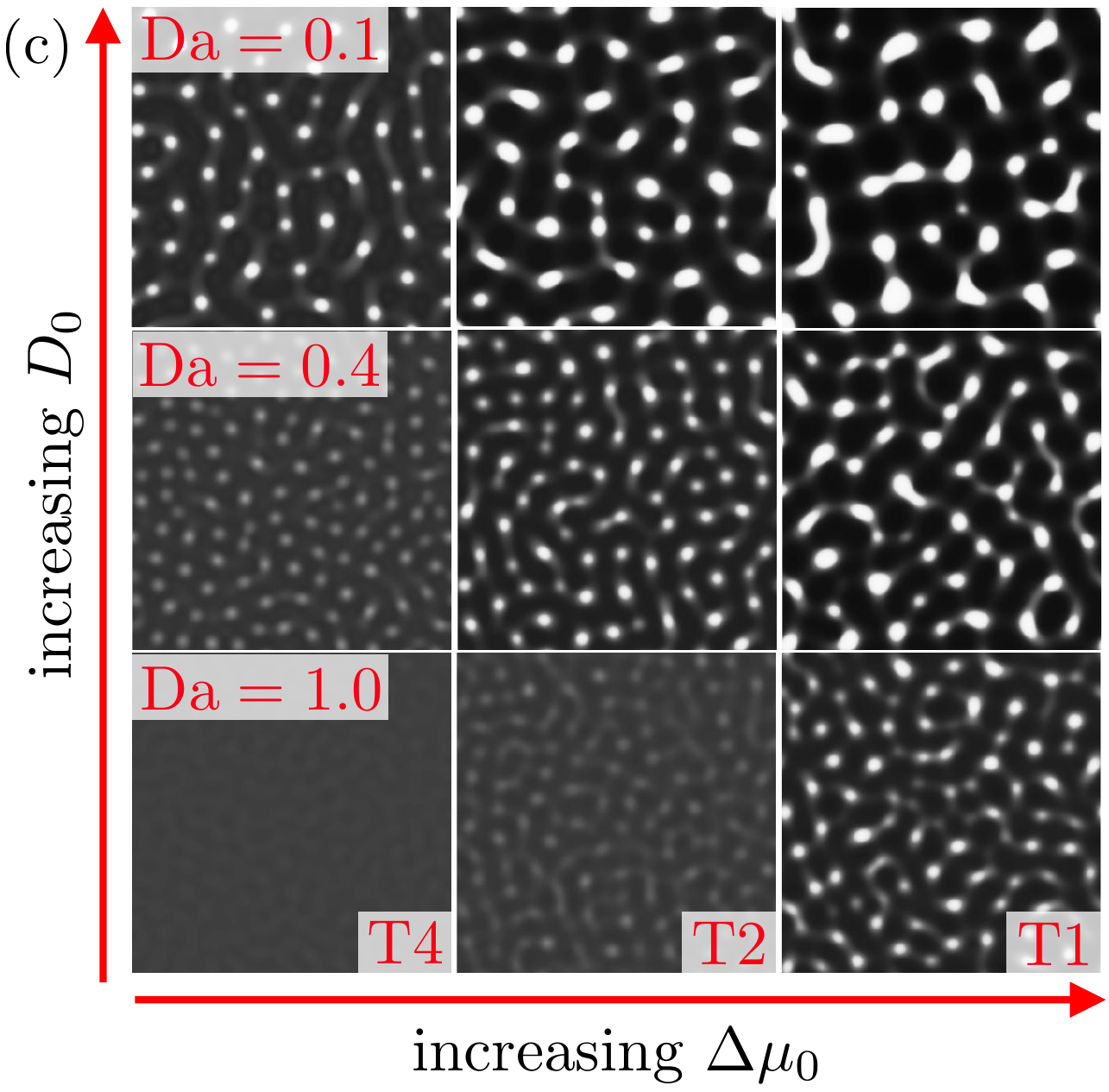}
\end{subfigure}
\caption{Snapshots of reactive systems for varying precursor landscapes $\mathrm{T}_i$ and $\mathrm{Da}$ at (a) $\tilde{\Phi} = 0.4$, (b) $\tilde{\Phi} = 0.6$, and (c) $\tilde{\Phi}=0.8$.}
\label{fig:4}
\end{figure*}

Our reaction rate $R$ adapts to evidence that many chemical systems demonstrate pathways to nucleation from stable prenucleation clusters~\cite{carcouet2014nucleation,banfield2000aggregation,gebauer2008stable,tan2014visualizing,zhang2007does,privman1999mechanism}. These intermediates accelerate (decelerate) nucleation by decreasing (increasing) the change in free energy $\Delta \mu = \mu-\mu_\mathrm{res}$ upon being added to the bulk structure~\cite{baumgartner2013nucleation}. Specifically, we denote $\Delta \mu$ as the difference in Gibbs free energy between a cluster precipitated into the local volume, $\mu$, and its constituent monomers dissolved in the external reservoir, $\mu_\mathrm{res}$: It is the diffusional chemical potential of a cluster in an open system~\cite{bazant2012theory}. In the following, we demonstrate nucleation of precursors that are stabilized by long-range electrostatic forces as is seen in, for instance, the early-stages of cement paste setting~\cite{ioannidou2016crucial,pellenq2004does} and sedimentation of charged nanoparticles~\cite{zhang2012non}, though we remark that the generality of the framework is readily adapted to alternative stabilization mechanisms.

For weakly screened particles with a surface charge, the classical energy contributions that scale with cluster volume and surface area are supplemented by a higher-order term that results from Coulomb interactions~\cite{groenewold2001anomalously}. Gibbs free energy of formation of such a cluster of $N $ monomers is approximated by~\cite{groenewold2001anomalously}
\begin{equation}
    \Delta \tilde{\mu}_\mathrm{c} = \tilde{\mu}_0(N)-\tilde{\mu}_\mathrm{res}= N \Delta \tilde{\mu}_0^\mathrm{m} + N^{2/3} \Lambda \tilde{\sigma} + N^{5/3} \frac{3}{2a}\lambda_\mathrm{B} \rho_\mathrm{c}^2 v^2,
    \label{eq:delta_mu}
\end{equation}
where $\Delta \tilde{\mu}_0^\mathrm{m}$ is the change in energy upon placing a free monomer into the interior of the cluster normalized by $k_\mathrm{B}T$, $\tilde{\sigma}$ is the surface energy, and $\Lambda$ is a constant shape factor. The electrostatic self-energy due to the particles' surface charge scales with the Bjerrum length $\lambda_\mathrm{B}$ and the square of the charge density $\rho_\mathrm{c}^2$~\footnote{Though $\rho_\mathrm{c}^2$ is inversely related to $\tilde{\phi}$ in closed systems, its dependence is less certain in open systems where ion exchange is possible and we hold it constant here}, and $v$ is the characteristic volume of a monomer. The $N^{5/3}$ dependence of this term derives from the number of Coulomb interactions measuring $\sim N^2$, while their mean separation distance measures $\sim N^{1/3}$. This choice of the free energy landscape, displayed in Fig.~\ref{fig:3}(a), allows a local minimum to form stable clusters en route to bulk nucleation. The free energy of formation of such a stable cluster is denoted by $\Delta \tilde{\mu}_0 = \Delta \tilde{\mu}_\mathrm{c}(N\simeq \pi a^3/6)$.

In a consistent description of reaction kinetics of nonequilibrium thermodynamic mixtures, the reaction complex explores the excess chemical potential landscape between local minima of cluster vacant sites $\mu_\circ^{\mathrm{ex}}$ and cluster occupied sites $\mu_\bullet^{\mathrm{ex}}$. If expressions for $\mu_\circ^{\mathrm{ex}}$ and $\mu_\bullet^{\mathrm{ex}}$ are known and the activation barrier of the transition state $\mu^\mathrm{ex}_\ddagger$ can be estimated, the net reaction rate is calculated from the probabilities of precipitating and dissolving stable clusters from a lattice as follows~\cite{bazant2012theory}:
\begin{widetext}
\begin{equation}
\label{eq:reaction_rate_1}
    R = k_0\left((1-\tilde{\phi})\exp\left(-\frac{\mu_\ddagger^\mathrm{ex}-\mu_\circ^\mathrm{ex}}{k_\mathrm{B}T}\right) - \tilde{\phi}\exp\left(-\frac{\mu_\ddagger^{\mathrm{ex}}-\mu_\bullet^\mathrm{ex}}{k_\mathrm{B}T}\right)\right),
\end{equation}
\end{widetext}
where the likelihood of cluster insertion or deletion is scaled by the fraction of vacant or occupied sites, respectively, and the attempt frequency, denoted by $k_0$, is assumed equal in both directions.

With reference to Appendix~\ref{app:chem_pot}, a thermodynamically consistent choice for the homogeneous parts of the chemical potentials of cluster occupied and cluster vacant sites read
\begin{subequations}
    \begin{align}
    \label{eq:vacancy}
        \mu_\circ^\mathrm{h} &= k_\mathrm{B}T\ln(1-\tilde{\phi}) + \mu_\circ^\mathrm{ex} = k_\mathrm{B}T\ln(1-\tilde{\phi})+\Omega\tilde{\phi}^2 + \mu_\mathrm{res}\\
    \label{eq:occupancy}
        \mu_\bullet^\mathrm{h} &= k_\mathrm{B}T\ln(\tilde{\phi}) + \mu_\bullet^\mathrm{ex} = k_\mathrm{B}T\ln(\tilde{\phi}) + \Omega(\tilde{\phi}-2)\tilde{\phi} + \mu_0,
    \end{align}
\end{subequations}
where $\mu_\circ^\mathrm{ex}$ is set to the chemical potential of the reservoir plus the mean change in interaction energy upon adding a vacancy, $\Omega \tilde{\phi}^2$, and $\mu_\bullet^\mathrm{ex}$ is set to the chemical potential of a cluster in a dilute solution plus the mean change in interaction energy upon inserting a cluster, $\Omega\tilde{\phi}(\tilde{\phi}-2)$. It is readily observed that the extensive property of the chemical potentials recovers Gibbs energy density, $g^\mathrm{h}=\phi \mu_\bullet^\mathrm{h} + (1-\tilde{\phi})\mu_\circ^\mathrm{h}$, and that replacing a vacancy by an occupancy is equivalent to inserting a particle from an external source, $\Delta \mu = \mu_\bullet - \mu_\circ$. If the transition state excludes one site during precipitation and dissolution reactions~\cite{bazant2012theory}, and we further estimate its excess chemical potential from the landscape of Gibbs energy of formation of a cluster in Eq.(\ref{eq:delta_mu}), we can write
\begin{equation}
\label{eq:transition_excess}
    \mu_\ddagger^\mathrm{ex} = -k_\mathrm{B}T \ln(1-\tilde{\phi}) + \alpha \mu_\bullet^{\mathrm{ex}} + (1-\alpha)\mu_\circ^\mathrm{ex} + \Delta \mu_\ddagger,
\end{equation}
where $\alpha$ is a symmetry factor measuring the fractional aggregation of monomers required to reach the transition state~\cite{kuznetsov1999electron}, and $\Delta \mu_\mathrm{cr} = \alpha \Delta \mu_0 + \Delta \mu_\ddagger=\mu_0^\ddagger - \mu_\mathrm{res}$ is the energy barrier with respect to the reservoir potential --- otherwise termed the energy of formation of a critical-sized cluster, for which a geometric interpretation is given in Fig.\ref{fig:3}(a). Hence, Eq.(\ref{eq:transition_excess}) assumes the excess chemical potential of the transition state, $\mu_\ddagger^{\mathrm{ex}}$, to be estimated from a weighted average of the excess chemical potentials of cluster inserted and cluster vacant states that are separated by a barrier of average height $\Delta \mu_\ddagger$; lastly, $k_\mathrm{B}T$ measures the energetic cost to the system in occupying a site during the transition. With the help of expressions in Eqn.(\ref{eq:vacancy}), (\ref{eq:occupancy}), and (\ref{eq:transition_excess}), the reaction rate in Eq.(\ref{eq:reaction_rate_1}) can be rewritten as a nonlinear function of $\Delta \mu$,
\begin{equation}
    \label{eq:reaction_rate}
    R =  R_0\left(\exp\left(-\alpha\frac{\Delta \mu}{k_\mathrm{B}T} \right) - \exp\left((1-\alpha)\frac{\Delta \mu}{k_\mathrm{B}T} \right)\right),
\end{equation}
for which the reaction rate coefficient obeys
\begin{equation}
\label{eq:reaction_rate_coef}
    R_0 = k_0 \tilde{\phi}^{\alpha}(1-\tilde{\phi})^{2-\alpha}\exp\left(-\frac{\Delta \mu_\ddagger}{k_\mathrm{B}T}\right).
\end{equation}
Eq.(\ref{eq:reaction_rate}) is the Allen-Cahn-Reaction rate for colloidal cluster formation, for which we highlight several salient features: i) It adheres to the De Donder relation requiring reaction rates to proceed in the direction of the chemical affinity, $\mathrm{sgn}(R)=-\Delta\mu$. ii) In the limit $\tilde{\phi}\to 0$, it recovers the dilute cluster nucleation rate $\dot{n}_0 = k_0 \exp(-\Delta \tilde{\mu}_\mathrm{cr})$, an expression consistent with classical nucleation theory. iii) The reaction rate coefficient $R_0$, adjusts the auto-catalytic behavior of $R$ through its dependence on the filling fraction, $\tilde{\phi}$.

At this point, it is helpful to recapitulate how physics at both the scale of the cluster and the mesoscale inform $R$. The formation barrier and size of a cluster is dictated by $\Delta \mu_\mathrm{c}(N)$ in Eq.(\ref{eq:delta_mu}) and sets the baseline relation between internal and external potentials $\Delta \mu_0$, as graphically represented in Fig.~\ref{fig:1}(c). Clusters form more readily if $\Delta \mu_0^\mathrm{m}$ decreases, the surface energy $\tilde{\sigma}$ reduces, or the electrostatic repulsion subsides. Importantly, clusters with attractive interactions can be metastable with respect to the reservoir potential, where $\Delta \mu_0 > 0$, yet lead to stable bulk nuclei at the mesoscale. In theory, the energy barrier to nucleate a high-density bulk phase from stable clusters is made up of the stabilizing inter-cluster repulsive forces and the kinetic energy of the clusters that transition from translational motion in the gas phase to vibrational motion in the jammed phase. In practice, the free energy density is fit to experimental observations of the miscibility gap and solubility limits by adjusting $\tilde{\Omega}$ and $\tilde{\kappa}_0$~\cite{cogswell2012coherency}; these quantities are otherwise difficult to measure. As a result, diluting effects of the electrostatic repulsion upon crowding of clusters and size-dependent scaling of the cluster-cluster interaction energy are subsumed into these mesoscopic parameters. Bulk nuclei form once the chemical driving force $\Delta \mu$ is sufficient to drive $\tilde{\phi}$ into the spinodal region, where Cahn-Hilliard dynamics promote phase separation under permitting characteristics of the reaction rate. These characteristics are explored next.

\begin{figure}
    \centering
    \includegraphics[width=0.4\textwidth]{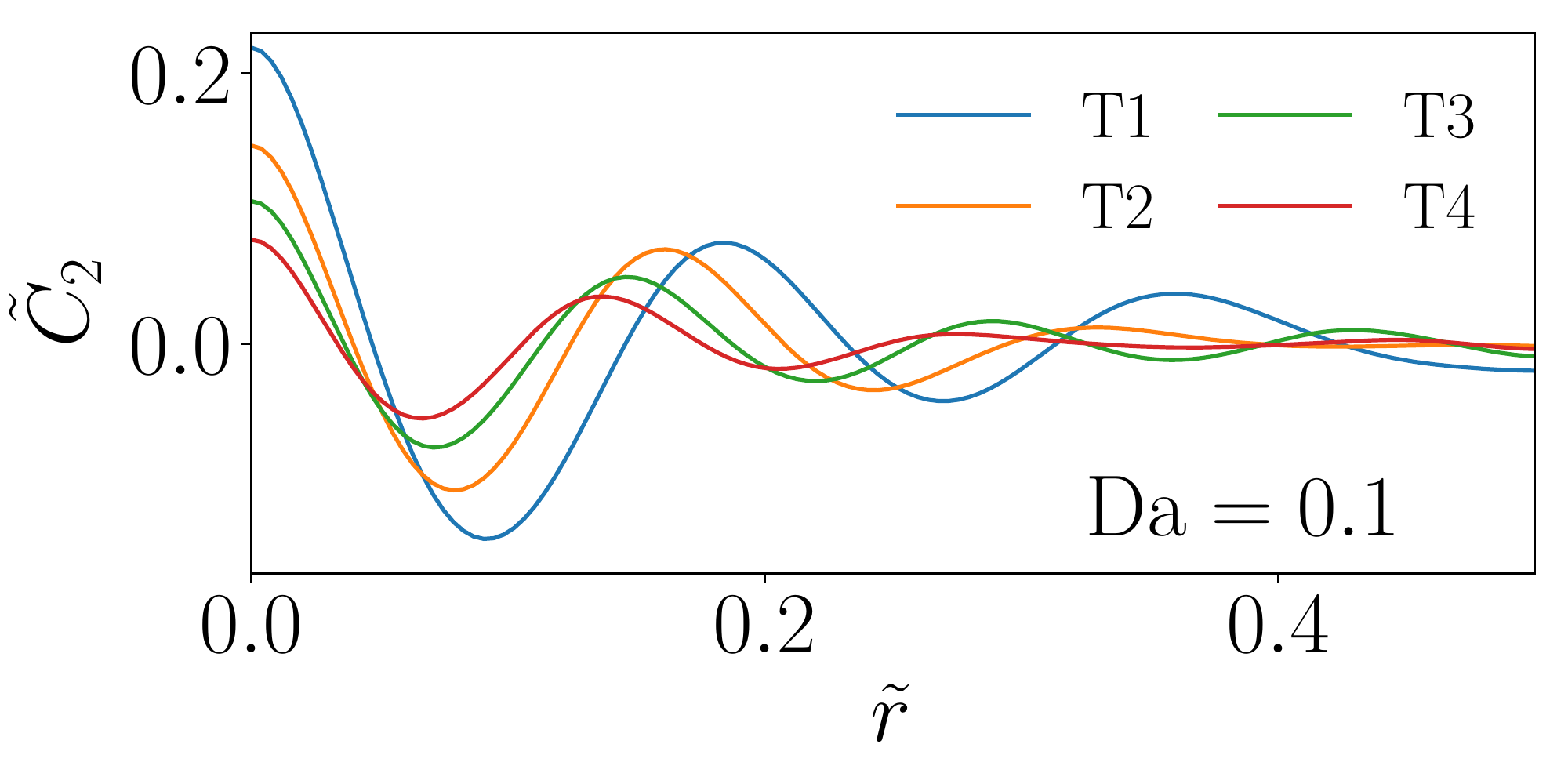}
    \caption{The two-point density correlation functions for $\mathrm{Da}=0.1$ as measured by the inverse Fourier transform of $S_q$ for $\tilde{\Phi}=0.6$ and $\tilde{\Omega}=4.0$.}
    \label{fig:two_point}
\end{figure}

\subsection{Results for nonconserved systems}

In this section, we explore pattern formation in reactive-diffusive systems by adjusting the thermodynamic landscape of clusters, while maintaining a constant reservoir potential and inter-cluster attraction. We elect to modify the landscape as follows: Decreasing $\Delta \mu_0^\mathrm{m}$ lowers $\Delta \mu_0$ and shifts the symmetry of the reaction rate toward lower $\alpha$ (see Fig.~\ref{fig:3}(a)). Fig.~\ref{fig:3}(b) plots four sample reaction rates, whose parameters are listed in Table~\ref{tab:params}. To make predictions about stability and dominant length scales, a linear stability analysis of the reaction-diffusion equation, Eq.(\ref{eq:reac_diff}), is performed in Appendix~\ref{app:linear_stability}. We show that stability is ensured by a negative growth rate coefficient, $\omega(\tilde{\phi}^\mathrm{h},q) < 0$, which measures the rate of change of small sinusoidal fluctuations $\delta \tilde{\phi}$ with wavenumber $q$ from a homogeneous base-state $\tilde{\phi}= \tilde{\phi}^\mathrm{h}$. For a purely reactive system, where $D = 0$, the auto-catalytic rate $\mathcal{A}$ predicts stability if~\cite{bazant2017thermodynamic},
\begin{equation}
    \mathcal{A}|_{\mu_\mathrm{res}}=\left(\frac{\partial R }{\partial \tilde{\phi}} + \frac{\partial R }{\partial \mu} \frac{\mathrm{d}\mu^\mathrm{h}}{\mathrm{d}\tilde{\phi}}\right) < 0
\end{equation}
The auto-catalytic rates are plotted on the secondary axis in Fig.~\ref{fig:3}(b), where it is shown that decreasing $\Delta \mu_0$ makes $R$ more auto-inhibitory. Thus, increasingly stable clusters suppress preferential precipitation onto existing bulk phases. To further analyze the implications of an auto-inhibitory reaction rate, Fig.~\ref{fig:3}(c) displays the integrated growth rate of unstable modes in course of reaction $\int \tilde{\omega} R \,\mathrm{d}\tilde{t}$, where $\tilde{\omega}$ is the linearized growth rate of the $\tilde{q}$ Fourier mode~\cite{bazant2017thermodynamic}. Growth rates are plotted for different $\Delta \mu_0$ and Damk\"{o}hler number $\mathrm{Da}=\dot{n}_0 l^2/D_0$, where $D_0=k_\mathrm{B}T/3\pi \eta a$ and $\dot{n}_0 = R(\tilde{\phi}\to 0)=k_0\exp(-\Delta \tilde{\mu}_\mathrm{cr})$ are the diffusivity and cluster nucleation rate in a dilute suspension, respectively. Increasing $\mathrm{Da}$ dampens the peak in $\int \tilde{\omega} R \,\mathrm{d}\tilde{t}$ caused by the Cahn-Hilliard kernel, and simultaneously decreasing $\Delta \mu_0$ --- that is, increasing the stability of the pre-nucleation clusters --- allows near complete suppression of mode growth.
\begin{figure}
\begin{subfigure}[b]{0.47\textwidth}
\includegraphics[width=1.0\textwidth]{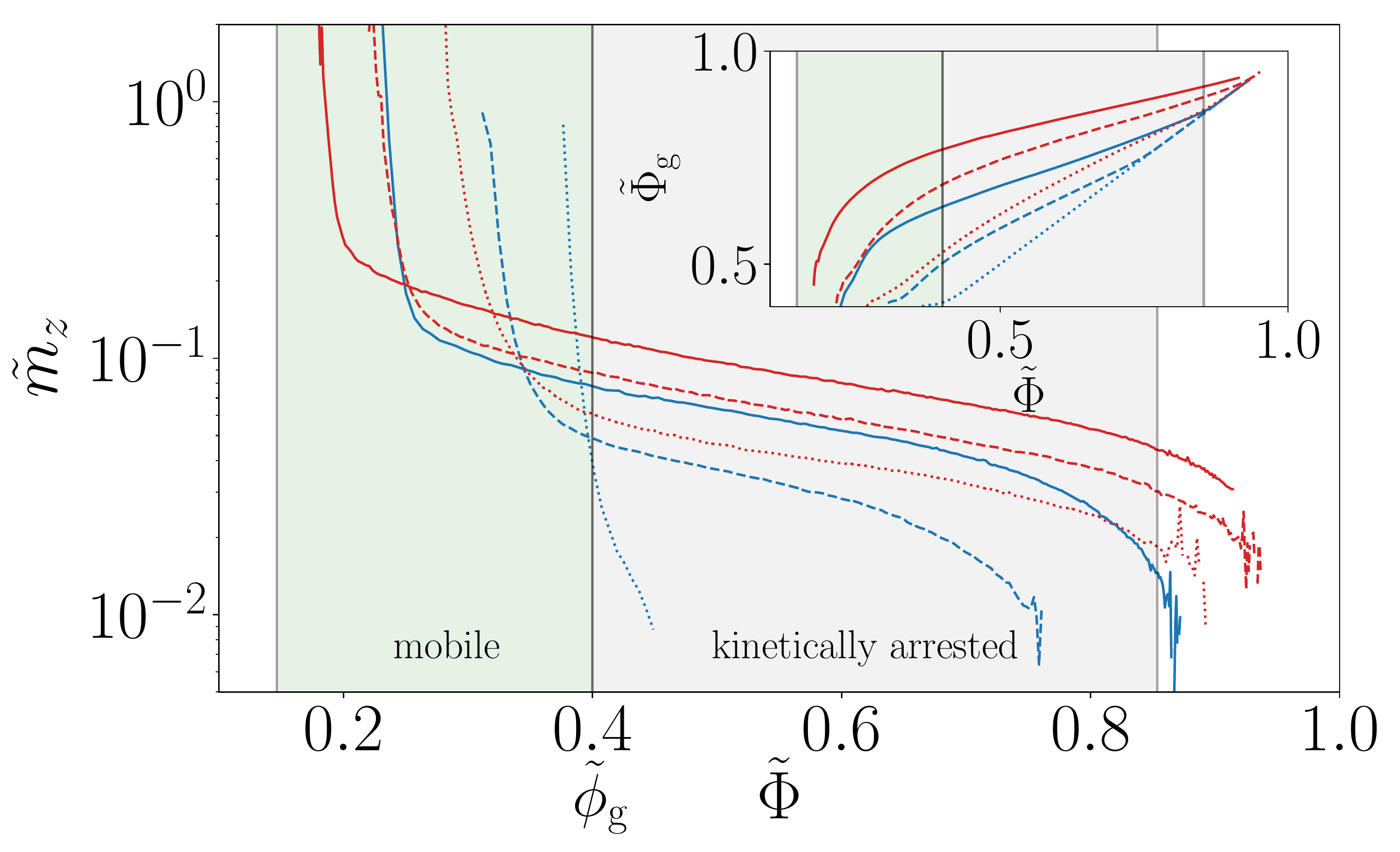}
\end{subfigure}
\caption{Mean pore size $\tilde{m}_z$ in function of the overall packing fraction. The shaded domain corresponds to the spinodal region, red and blue lines correspond to T1 and T4, respectively, and solid, striped, and dotted lines correspond to $\mathrm{Da}=0.1$, $0.4$., and $1.0$, respectively.}
\label{fig:5}
\end{figure}

These predictions are verified by the pattern evolution depicted in Fig.~\ref{fig:4}(a)-(c) for increasing reaction extents $\tilde{\Phi} = \{0.4,0.6,0.8\}$. The characteristic size of the emergent bulk nucleii scale as $\sim(\kappa/(\mathrm{d}\mu^\mathrm{h}/\mathrm{d}\tilde{\phi}))^{1/2}$ and depend critically on the location within the spinodal region at which fluctuations grow. Strongly diffusive clusters (low $\mathrm{Da}$) phase separate readily near the spinodal, amplifying large wavelengths, whereas reaction-controlled dynamics (high $\mathrm{Da}$) delay the growth of instabilities toward higher $\tilde{\Phi}$, forming smaller nucleii in larger abundance or proceeding by spinodal decomposition. Importantly, dynamic arrest in diffusion at $\tilde{\phi}_\mathrm{g}$ halts Cahn-Hilliard mode growth, indicating that microtextural patterns in attractive colloids are determined in the low density portion of the spinodal region. As $\Delta \mu_0$ is decreased, further cluster insertion shifts from surface growth of bulk nucleii toward homogeneous densification; similar observations were made in molecular-dynamics simulations of crystallizing Lennard-Jones fluids in Ref.~\cite{trudu2006freezing}. These physics are further reflected in the two-point density correlation function $\tilde{C}_2$ plotted in Fig.~\ref{fig:two_point}(b), where auto-inhibitory kinetics moderate its amplitude and move the location of the second peak toward closer separation distances $\tilde{r}$. That is, as $\mathcal{A}$ becomes less positive, the mean packing fraction of the gel-phase $\tilde{\Phi}_\mathrm{g}$ behaves increasingly linearly with the overall reaction extent (see inset of Fig~\ref{fig:5}) and density fluctuations form at smaller wavelengths. Lastly, we calculate the mean pore size $m_z$ from the pore-chord length probability density function with threshold $\tilde{\phi}_\mathrm{g}$ between gas and gel phases, using the following relation~\cite{torquato2013random}:
\begin{equation}
    m_z = \int_0^\infty z p(z)\mathrm{d}z,
\end{equation}
where $p(z)$ is the pore-chord length probability density function and $z$ measures the length of a sampled chord. As seen in Fig.~\ref{fig:5}, the location of onset of bulk nucleation predicts $\tilde{m}_z$ prior to entering Avrami-like growth, where precipitation at lower $\Delta \mu_0$ universally produces smaller pores. This reduction in pore size is facilitated both by accessing smaller wavelengths at phase separation, and more uniformly precipitating stable precursors once the cluster aggregates have dynamically arrested. Mesoscale texture is critical in predicting a host of important material properties --- elasticity, fracture toughness, fluid and electrical conductivity, to name a few --- and Fig.~\ref{fig:5} shows that pattern formation can be controlled if the reaction rate is adjusted within the mobile portion of the spinodal region. The most important finding of our model is that the thermodynamic landscape of the prenucleation clusters is instrumental in suppressing or enhancing mesoscopic: By increasing the stablility of the clusters, which concurrently lowers $\alpha$, the range of the unstable growth modes of the Allen-Cahn reaction kernel narrows and shifts outside the spinodal region of the Cahn-Hilliard diffusion kernel.
\section{Conclusions}
In summary, we have developed a mean-field, nonequilibrium thermodynamic model for interacting colloids that respects experimental observations made at two essential length scales: Clusters nucleate as stable building blocks at the microscale~\cite{weitz1984fractal}, and aggregate into an arrested, out-of-equilibrium structure at the mesoscale~\cite{lu2008gelation,ioannidou2016mesoscale}. Including a variable, entropic gradient energy penalty in Gibbs free energy provides a new physics-grounded approach to simulate the evolution of mean-field gel patterns. Previous studies have proposed control of the reaction rate to modulate dominant wavelengths during phase separation~\cite{glotzer1995reaction,bai2011suppression,bazant2017thermodynamic}. But the expressions derived in Eq.(\ref{eq:reaction_rate}) and Eq.(\ref{eq:reaction_rate_coef}), which marry the reactive landscape of stable intermediates into the Allen-Cahn-Reaction equation~\cite{bazant2012theory}, lend new precision: They provide a template to manipulate bulk colloidal structures that emerge from stable pre-nucleation clusters by adjusting $\Delta \mu_0$ upon entering the mobile portion the spinodal region. While, the present study focused on the response of homogeneous systems to sudden changes in the normalized attractive strength and supersaturation of the solution (i.e., $\Delta \tilde{\mu}_0^\mathrm{m}$), future work should assess how adjustments in the reservoir potential in course of reaction can manipulate ensuing colloidal patterns. This could advance design of tailor-made colloidal structures that enhance selected mechanical properties.

\section*{Acknowledgements}
The authors would like to thank Amir Pahlavan and Thibaut Divoux for insightful discussions and references. Financial support was provided by the National Science Foundation Graduate Research Fellowship, and the Concrete Sustainability Hub at the Massachusetts Institute of Technology (CSHub@MIT) with sponsorship provided by the Portland Cement Association (PCA) and the Ready Mixed Concrete (RMC) Research and Education Foundation. Additional support was provided by the ICoME2 Labex (ANR-11-LABX-0053) and the A*MIDEX projects (ANR-11-IDEX-0001-02) cofunded by the French program ``Investissements d'Avenir" managed by the French National Research Agency.
\\
\appendix
\begin{widetext}
\section{Chemical potential of vacancies and an occupancies}
We briefly outline relations for the homogeneous chemical potentials of a vacancy $\mu_\circ^\mathrm{h}$ and an occupancy $\mu_\bullet^\mathrm{h}$. Using the classical thermodynamic definition of a chemical potential, $\mu_\circ^\mathrm{h}$ quantifies the change in Gibbs free energy due to a change in the number of vacancies (voids) $N_\circ$ while keeping the temperature $T$, the pressure $P$ (which we have not explicitly defined), and number of occupancies (clusters) $N_\bullet$ constant,
\label{app:chem_pot}
\begin{subequations}
    \label{eq:chem_pot_thermo}
    \begin{align}
        \mu_\circ &= \left(\frac{\partial G}{\partial N_\circ}\right)_{N_\bullet,T} =\left(\frac{\partial G}{\partial V}\right)_{\phi,T}\left(\frac{\partial V}{\partial N_\circ}\right)_{N_\bullet,T} + \left(\frac{\partial G}{\partial \tilde{\phi}}\right)_{V,T} \left(\frac{\partial \tilde{\phi}}{\partial N_\circ}\right)_{N_\bullet,T}.
    \end{align}
\end{subequations}
Above, $V$ is the system volume and an equivalent expression can be written for $\mu_\bullet$ by replacing $N_\circ$ with $N_\bullet$.  Assuming $G = V[n_\mathrm{s}k_\mathrm{B}T(\tilde{\phi}\ln(\tilde{\phi})+(1-\tilde{\phi})\ln(1-\tilde{\phi})-\tilde{\phi}^2\tilde{\Omega})+\tilde{\phi}\mu_0 + (1-\tilde{\phi})\mu_\mathrm{res}]$ and noting the total number of sites as $N_\mathrm{tot} = V n_\mathrm{s} = N_\circ + N_\bullet$ with packing density $\tilde{\phi} = N_\bullet / N_\mathrm{tot}$, Eq.(\ref{eq:chem_pot_thermo}) readily yields
\begin{subequations}
    \begin{align}
        \mu_\circ^{\mathrm{h}} & = (g^\mathrm{h}/n_\mathrm{s}) - \phi \mu^\mathrm{h} = k_\mathrm{B}T \ln(1-\tilde{\phi}) + \Omega \tilde{\phi}^2 + \mu_\mathrm{res} \\
        \mu_\bullet^\mathrm{h} & = (g^\mathrm{h}/n_\mathrm{s}) + (1-\phi) \mu^\mathrm{h} = k_\mathrm{B}T\ln(\tilde{\phi}) + \Omega (\tilde{\phi}-2)\tilde{\phi} + \mu_0.
    \end{align}
\end{subequations}
Here, $\Omega \tilde{\phi}^2$ is the mean change in interaction energy attributed to adding a new site into the lattice, while $-2\Omega\tilde{\phi}$ measures the mean change in interaction energy upon placing a cluster into that site. The derivation for the inhomogeneous case may similarly be derived using the calculus of variations.

\section{Linear stability analysis of the reaction-diffusion equation}
\label{app:linear_stability}

This Appendix outlines the criterion for stability of our reaction-diffusion equation that describes the nonequilibrium thermodynamics of attractive colloids, Eq.(\ref{eq:reac_diff}). The derivation follows closely the procedure outlined by one of the author in Ref.~\cite{bazant2017thermodynamic}, which may be consulted for additional details. Starting from a homogeneous base state where $\tilde{\phi}(\mathbf{x})=\tilde{\phi}^\mathrm{h}$, the field is perturbed by small fluctuations $\delta \tilde{\phi}$. We measure the aggregate strength of the fluctuations by the $l^2$-norm of the perturbation field,
\begin{equation}
    \mathcal{L} = \frac{1}{2}\int_V (\delta \tilde{\phi})^2\mathrm{d}V.
\end{equation}
For a base state to be stable with respect to the dynamics imposed by Eq.(\ref{eq:reac_diff}), $\mathcal{L}$ must be a decreasing function of time. That is,
\begin{equation}
\label{eq:Lyapunov}
    \text{Stable if: }\qquad \frac{\mathrm{d}\mathcal{L}}{\mathrm{d}t} = \int_V \delta \tilde{\phi}\left(\frac{\partial \delta \tilde{\phi}}{\partial t}\right)\mathrm{d}V = \int_V \left(-\mathcal{D}(\nabla \delta \tilde{\phi})^2 + \mathcal{A}(\delta \tilde{\phi})^2\right)\mathrm{d}V < 0,
\end{equation}
where the chemical diffusion $\mathcal{D}$ and auto-catalytic rate $\mathcal{A}$,
\begin{subequations}
\begin{align}
    \mathcal{D} &= L\frac{\delta \tilde{\mu}}{\delta \tilde{\phi}} \\
    \mathcal{A} &= \frac{\delta R}{\delta \tilde{\phi}} = \frac{\partial R}{\partial \tilde{\phi}} + \frac{\partial R}{\partial \tilde{\mu}}\frac{\delta \tilde{\mu}}{\delta \tilde{\phi}} + \frac{\partial R}{\partial \tilde{\mu}_\mathrm{res}} \frac{\partial \tilde{\mu}_\mathrm{res}}{\partial \tilde{\phi}},
\end{align}
\end{subequations}
are evaluated at $\tilde{\phi}(\mathbf{x})=\tilde{\phi}^\mathrm{h}$ and $L=D\phi/k_\mathrm{B}T$ is the Onsager coefficient. Next, the variational derivative of the internal chemical potential $\tilde{\mu}$ simplifies to
\begin{equation}
    \frac{\delta \tilde{\mu}}{\delta \tilde{\phi}} = \frac{\delta^2 G}{n_\mathrm{s}k_\mathrm{B}T(\delta \tilde{\phi})^2} = \frac{\mathrm{d}\tilde{\mu}^\mathrm{h}}{\mathrm{d}\tilde{\phi}} + \tilde{\kappa} \frac{(\nabla \delta \tilde{\phi})^2}{(\delta \tilde{\phi})^2},
\end{equation}
once insignificant terms dependent on $\nabla\tilde{\phi}$ are removed. Lastly, we choose the perturbation to be sinusoidal with Fourier frequency $q$, for which $\nabla\delta \tilde{\phi} = q\delta\tilde{\phi}$. If the growth rate of the perturbation is linearized as $(\partial(\delta \tilde{\phi})/\partial t) = \omega \delta \tilde{\phi}$ with growth rate coefficient $\omega$, the stability criterion in Eq.(\ref{eq:Lyapunov}) can be rewritten as
\begin{equation}
    \text{Stable if: }\qquad \frac{\mathrm{d}\mathcal{L}}{\mathrm{d}t} = \omega \int_V (\delta \tilde{\phi})^2\mathrm{d}V < 0,
\end{equation}
with
\begin{equation}
\label{eq:grwoth_rate}
    \omega(\tilde{\phi}^\mathrm{h}, q) = \left(\frac{\partial R}{\partial \tilde{\phi}} + \frac{\partial R}{\partial \tilde{\mu}_\mathrm{res}} \frac{\partial \tilde{\mu}_\mathrm{res}}{\partial \tilde{\phi}}\right) + \left(\frac{\partial R}{\partial \tilde{\mu}}-L q^2\right)\left(\frac{\mathrm{d}\tilde{\mu}^\mathrm{h}}{\mathrm{d}\tilde{\phi}} + \tilde{\kappa} q^2\right).
\end{equation}

\end{widetext}

\bibliographystyle{aipnum4-1}
\bibliography{mybibliography}

\end{document}